\documentclass[a4paper,11pt]{article}
\usepackage{jinstpub} 
\usepackage{enumitem}
\usepackage[capitalize]{cleveref}

\graphicspath{{figs/}}

\title{AugerPrime Surface Detector Electronics}

\author{\includegraphics[height=30mm]{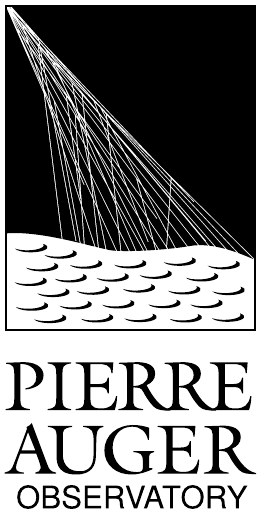}\\[3mm]The Pierre Auger Collaboration}
\affiliation{Av.\ San Mart\'{\i}n Norte 306, 5613 Malarg\"ue, Mendoza, Argentina}
\emailAdd{spokespersons@auger.org}

\abstract{Operating since 2004, the Pierre Auger Observatory has led to major advances in our understanding of the ultra-high-energy cosmic rays. The latest findings have revealed new insights that led to the upgrade of the Observatory, with the primary goal of obtaining information on the primary mass of the most energetic cosmic rays on a shower-by-shower basis.
In the framework of the upgrade, called AugerPrime, the 1660 water-Cherenkov detectors of the surface array are equipped with plastic scintillators and radio antennas, allowing us to enhance the composition sensitivity. To accommodate new detectors and to increase experimental capabilities, the electronics is also upgraded. This includes better timing with up-to-date GPS receivers, higher sampling frequency, increased dynamic range, and more powerful local processing of the data.
In this paper, the design characteristics of the new electronics and the enhanced dynamic range will be described. The manufacturing and test processes will be outlined and the test results will be discussed. The calibration of the SD detector and various performance parameters obtained from the analysis of the first commissioning data will also be presented.
}


\begin{document}
\maketitle

\section{Introduction}

The Pierre Auger Observatory is located near Malarg\"ue, Mendoza, Argentina. The surface detector (SD) array of the observatory consists of 1600 water-Cherenkov detectors (WCD) on a 1500\,m triangular grid covering 3000\,km$^2$. Another 60 WCDs, with a 750\,m spacing, form a 27\,km$^2$ infill region allowing extension to lower energies. The array is overlooked by four fluorescence detector (FD) sites each hosting 6 telescopes viewing a $180^\circ$ azimuth by $30^\circ$ elevation field of view.  Three additional telescopes at one of the sites can be tilted $30^\circ$ higher to view lower energy showers and overlook the infilled surface array.

Secondary particles of extensive air showers (EAS) induced by ultra-high-energy cosmic rays (UHECRs) are sampled at ground level by the SD. The FD measures EAS development by detecting the nitrogen UV light produced by the shower particles along their passage through the atmosphere. Additional instrumentation for R\&D on muon (UMD) and radio-based (AERA) detection is also located on the site. A description of the current observatory can be found in Ref.~\cite{NIM}.

In almost 20 years of operation, the Pierre Auger Observatory has provided, with unprecedented statistics and precision, major breakthroughs in the field of UHECRs. The steepening of their flux is now confirmed beyond any doubt as a succession of different power laws~\cite{Spectrum}. The primary mass composition is found to get heavier with increasing energy~\cite{Mass}. A large-scale anisotropy has been discovered above $8{\times}10^{18}$\,eV, proving that these UHECRs are of extragalactic origin~\cite{Dipole}, while anisotropies that mirror the distribution of nearby extragalactic matter have been evidenced at intermediate angular scales above ${\simeq}4{\times}10^{19}$\,eV~\cite{EGal}. Furthermore, important results have been obtained also for neutrinos and photons.

To make further progress, the Auger Collaboration decided to improve the SD sensitivity to the cosmic ray composition.  The Observatory is therefore undergoing a significant upgrade of its experimental capabilities called AugerPrime, with the main aim of disentangling the muonic and electromagnetic components of extensive air showers, thereby enhancing the ability to study UHECR composition.
This will allow us to understand the origin of the flux suppression, providing fundamental constraints on the sources and their properties, to perform composition-assisted anisotropies, and to add information about hadronic interaction effects at the highest energies. Enhanced trigger capabilities will furthermore provide higher sensitivity to neutrinos and photons.

After a brief description of the different components of AugerPrime, the design of the Surface Detector upgraded electronics will be described in the following. The test processes and the various test results will be presented. The calibration of SD stations will be outlined.  Finally, the detector performance inferred from the analysis of the first data will be discussed.

\section{AugerPrime components}

A WCD consists of a 3.6\,m diameter tank containing a sealed liner with a reflective inner surface. The liner contains 12\,000 liters of ultra-pure water. Three 9-inch diameter Photonis XP1805/D1 photomultiplier tubes (PMTs) are symmetrically distributed on the surface of the liner at a distance of 1.20\,m from the tank center axis and look downward through windows of clear polyethylene into the water. They record the Cherenkov light produced by the passage of relativistic charged particles through the water. The tank height of 1.2\,m makes it also sensitive to high
energy photons, which convert to electron-positron pairs in the water volume. Each surface detector station is self-contained. A solar power system provides currently an average of 10\,W for the PMTs and electronics package consisting of a processor, Global Positioning System (GPS) receiver, radio transceiver and power controller. 

To increase the dynamic range of the WCD signal measurement, a small PMT (SPMT), a 1-inch Hamamatsu R8619 PMT, dedicated to the unsaturated measurement of large signals, is added to one of WCD liner ports. An already existing LED flasher is mounted to another port of the water tank liner. The LED flasher incorporates two LEDs which can be pulsed independently or simultaneously and with variable amplitude. This allows testing of the linearity of the photomultipliers to be conducted remotely.

A scintillator-based surface detector (SSD) consists of an aluminum box of $3.8\,\text{m}\times1.3$\,m, containing two scintillator panels, each composed of extruded
polystyrene scintillator bars of 1.6\,m length, 5\,cm width, and 1\,cm thickness. The scintillator light is read out with wavelength-shifting fibers inserted into straight extruded holes in the scintillator bars. The 1-mm diameter fibers are bundled in a PMMA (poly(methyl methacrylate)) cylinder which is connected to a single PMT. The PMT is a 1.5-inch diameter bi-alkali Hamamatsu R9420. The power supply of the PMT is based on a custom design manufactured by the ISEG company. The charge value for a Minimum Ionizing Particle (MIP) determined by using a hodoscope trigger is more than 30 photo-electrons (p.e.). 

The Radio Detector (RD) is a short aperiodic loaded loop antenna of 122~cm diameter, measuring radio signals from extensive air showers in the 30 to 80\,MHz band. 
It features a simple mechanical design, minimizing cost and easing handling and maintenance. The antenna features a 392\,$\Omega$ resistor at the bottom, which shapes the antenna main lobe towards the zenith and suppresses the dependence on structures below it, in particular the SSD, the WCD and potentially variable ground conditions. The SSD and RD are mounted atop each WCD detector except for the detector stations on the border of the array where the shower core measurement is no longer necessary since a high-lever trigger requires a ring of stations around the shower core.

In addition, an Underground Muon Detector (UMD), consisting of buried muon counters deployed in the infill area, gives a direct measurement of the muon content of the showers and of its time structure. The UMD basic unit consists of $3\times10$\,m$^2$ modules, each segmented into 64 plastic scintillator strips, buried 2.3\,m alongside a WCD at a distance of at least 7\,m.

The upgrade of the SD electronics (SDEU) allows us to process signals from SSD and SPMT, in addition to those of the WCD large PMTs, to obtain an absolute time indication, and to provide digital interface for RD and UMD detectors. Furthermore, the new electronics is designed to improve both resolutions and data processing capabilities.  
In the main array, the existing communication infrastructure of the stations is used, and therefore, no upgrade of the main communication system is required. The station power system remains unchanged except for new solar panels to accommodate the increased power consumption due to the RD. 

\begin{figure}
\centering
\includegraphics[width=0.7\textwidth]{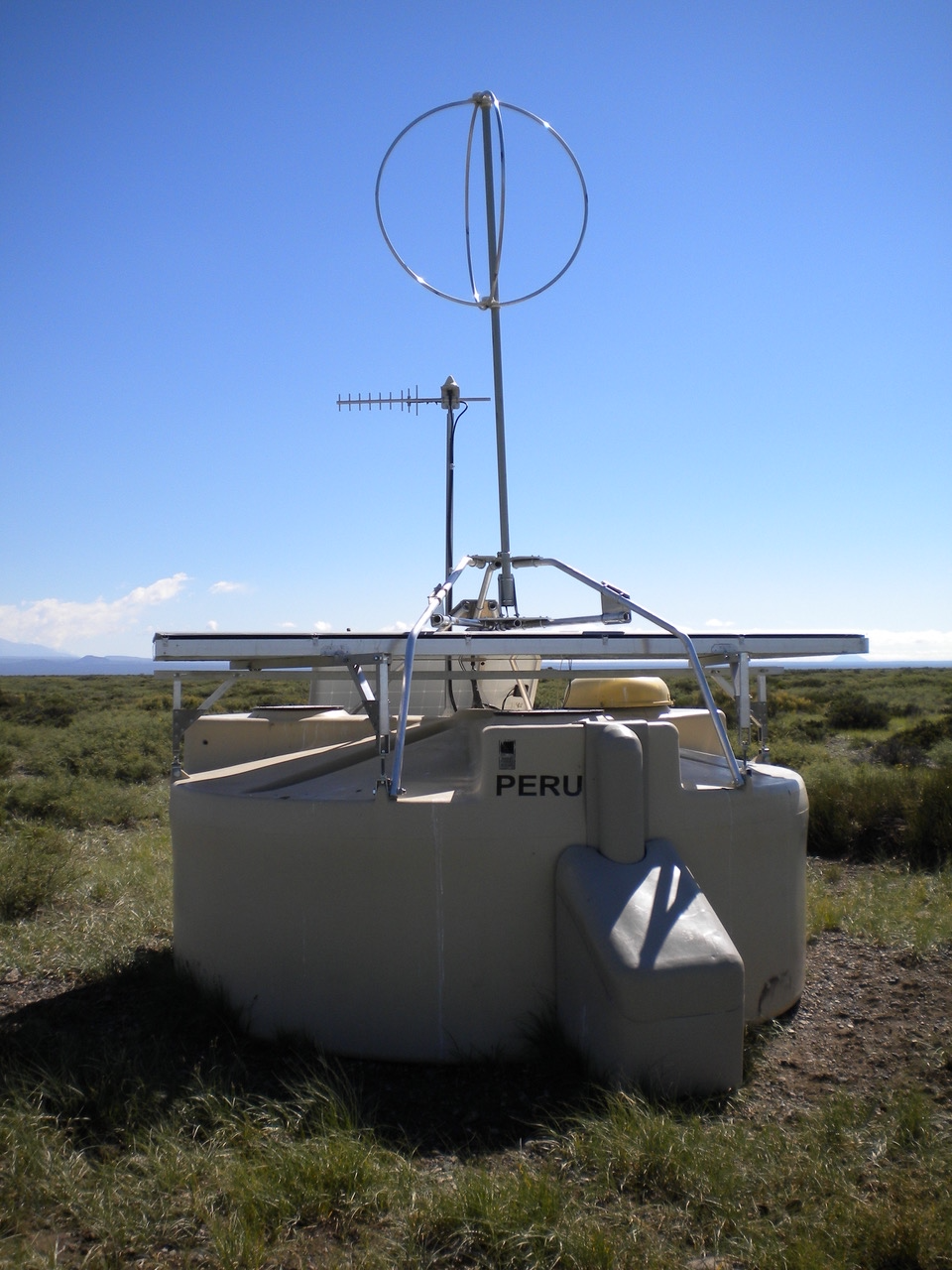}
\caption{AugerPrime detector with the SSD and RD atop the WCD. The UUB is hidden underneath the dome visible on top of the WCD.}
\label{fig:AugerPrimeDet}
\end{figure}

A general description of AugerPrime and its physics motivations can be found in the Preliminary Design Report~\cite{PDR}.
An AugerPrime detector station with the SSD scintillator and the RD antenna atop the WCD detector is shown in \cref{fig:AugerPrimeDet}. 

An AugerPrime engineering array (EA) of 12 stations has been operating at the Auger Observatory site since October, 2016. The EA allowed us to validate the design and to test the integration of the AugerPrime stations into the standard Observatory operation and the Central Data Acquisition System (CDAS) through the Auger communication network. The description of the preliminary design and the results obtained from the EA can be found in Refs.~\cite{Lagorio,Suomijarvi,Zong}.

The deployment of the pre-production and production electronics, together with SPMTs, started in mid-2020. All the PMTs are procured and tested and the production of the electronics boards is completed.  The deployment on-site was completed early July 2023. The commissioning studies have been in progress since December 2020 and various performance parameters are being monitored. 
The RD detector prototypes have been tested on the SD array; their installation started mid-2023 and it is planned to be completed by early 2024. The installation of UMD is well advanced too and is foreseen to be completed by early 2025.

\section{Requirements and general implementation}

The global design objectives of the electronics upgrade are to increase the data quality: faster sampling for Analog-Digital-Converter (ADC) traces, better timing accuracy, increased dynamic range, enhanced local trigger and processing capabilities, more powerful local station processor with a Field Programming Gate Array (FPGA), and improved calibration and monitoring capabilities.  Backwards-compatibility with the current dataset is maintained by retaining the current timespan of the PMT-traces and providing for digital filtering and downsampling of the traces to emulate the current triggers in addition to any new triggers.  The design objectives also aim for higher reliability and easy maintenance. 
The most important functional and configuration requirements are listed below followed by a description of the general implementation.

\subsection{Functional requirements}

\begin{itemize}
\item 10 ADC analog inputs to handle the two gains for each of the three existing PMTs, the added PMT of the SSD detector and the SPMT (plus a spare channel).
\item The total RMS integrated noise at the ADC input should not exceed 0.5\,LSB (Least Significant Bit) for the low-gain channel and 2\,LSB for the high-gain channel.
\item Digitization of the PMTs anode signals at a sampling frequency of 120\,MS/s with a resolution of 12\,bit minimum.
\item Existing and additional trigger configurations implemented in the FPGA firmware.
\item Event time tagging with a resolution of 5\,ns with a stability better than 5\% depending on temperature variation.
\item Independent programmable Slow-Control unit to monitor voltage and environmental sensors, and control the PMT high voltages and the FPGA low voltages.
\item Calibration system based on two LEDs, controlled in time and amplitude.
\item Ethernet and USB (Universal Serial Bus) communication capabilities.
\end{itemize}

\subsection{Configuration requirements}

\begin{itemize}
\item All functions contained on a single board (except for the GPS receiver).
\item Use of up to date commercial GPS receivers.
\item Embedded diagnostics.
\item Digital ports allowing communication with additional detector systems.
\item Power-supply unit including safety features and an efficiency better than 80\% for a total consumption between 10 and 11\,W.
\end{itemize}

\subsection{Electronics implementation}

The major portion of the AugerPrime electronics upgrade replaces the original Unified Board (UB) with the Upgraded Unified Board (UUB). In the UUB,  various functions (front-end, calibration, time tagging, trigger, monitoring) are implemented on a single board. It is designed to fit the existing RF-enclosure, and to accept the SSD PMT and SPMT cables together with the existing PMTs, GPS antenna, and communications cables. In addition, UUB provides digital interface for the RD and UMD detectors giving them access to the communication system. The new electronics also employs faster ADCs (120\,MHz instead of 40\,MHz) with larger dynamic range (12\,bit each instead of 10\,bit). 

The UUB architecture is designed with a Xilinx Zynq FPGA containing two embedded ARM Cortex  A9 333\,MHz microprocessors. The FPGA is connected to a 4\,Gbit LP-DDR2 memory and a 2\,Gbit Flash memory. The FPGA  implements all basic digital functions such as the read-out of the ADCs, the generation of triggers, the interface to LED flasher, GPS receiver, clock generator, and memories. High-level functions like the data handling and communications with the radio transceiver are implemented under Linux. 
A simplified functional diagram of the UUB architecture is shown in \cref{fig:Zynq}. A more detailed diagram is shown in \cref{fig:Zynq2} in the Appendix.

\begin{figure}
\centering
\includegraphics[height=0.9\textwidth,angle=-90]{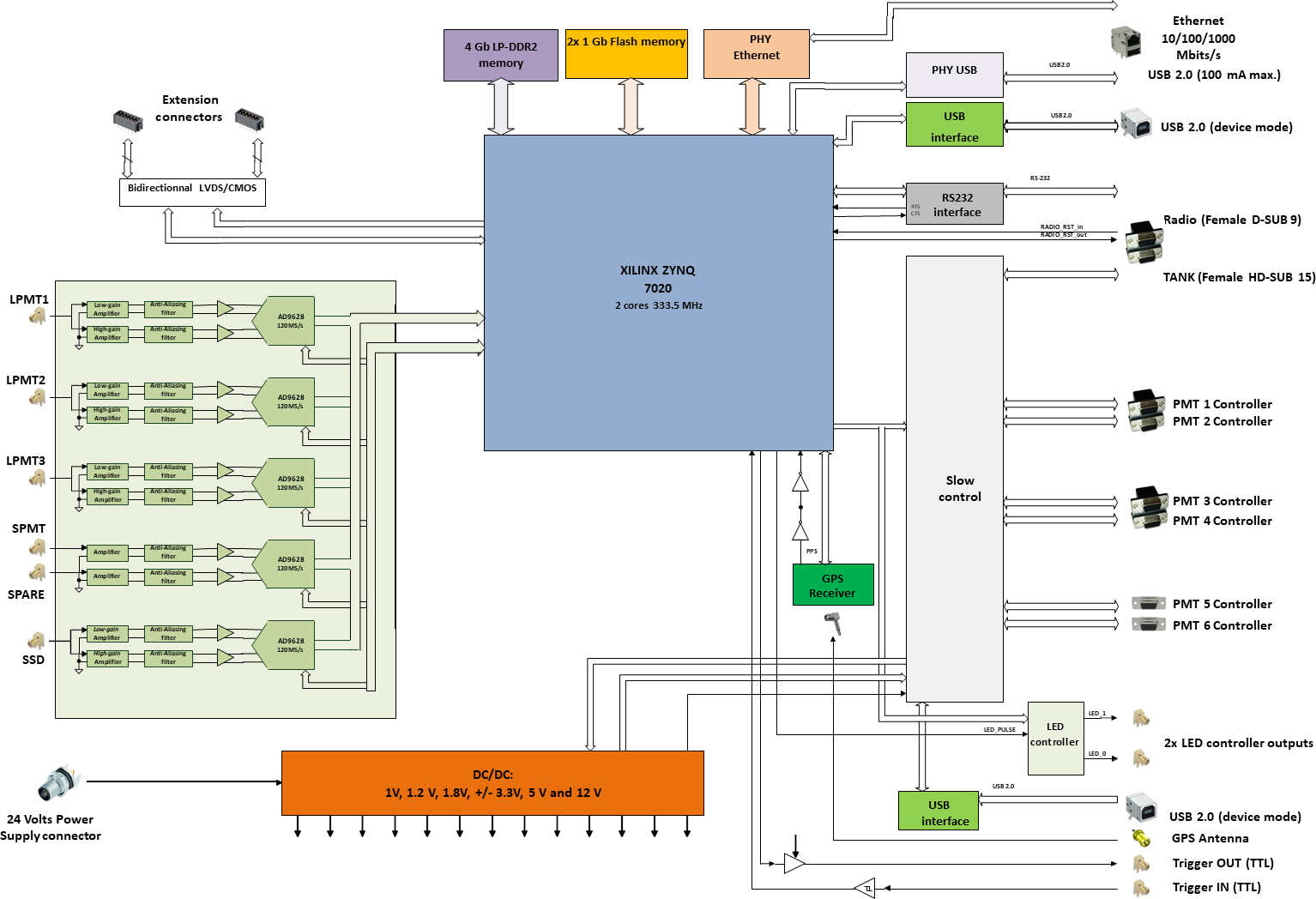}
\caption{Functional diagram of the Upgraded Unified Board.}
\label{fig:Zynq}
\end{figure}

\section{Design characteristics}

The station electronics was designed to use more advanced and less power consuming electronics components.  It takes advantage of the existing mechanical interfaces, and the existing communication and power systems.  Furthermore, the new firmware/software was adapted from the previous one ensuring compatibility with the Central Data Acquisition System (CDAS). 
In the following, the design characteristics for the different components of the AugerPrime Surface Detector electronics together with the added SPMT, are described.

\subsection{The Surface Detector dynamic range\label{subsec:dynamicrange}}

The dynamic range of the SD measurements extends from a few photoelectrons in stations far from the shower core and for the low energy muons used for calibration, to hundreds of thousands in stations near the impact point of the shower core at the ground where the particle density dramatically increases.
To improve the SD data quality, an extension of the acquisition dynamic range is implemented in both the WCD and the SSD, allowing us to measure non-saturated signals at distances as close as 250\,m from the shower core, in particular for the highest energy events, which are of extreme importance for the physics goals. 

To achieve this aim, the WCD is equipped with an additional small PMT (SPMT), a 1-inch Hamamatsu R8619 photomultiplier, assembled with a pure passive 66.5\,M$\Omega$ tapered ratio HV divider for high linearity and low power consumption. The SPMT is installed in an hitherto unused and easily accessible 30\,mm window on the Tyvek bag containing the ultra-pure water, located close to one of the large PMTs (LPMT1). 
The SPMT features the same bialkali photocathode as the XP1805 LPMTs, but with an active area of about 1/80, thus potentially allowing
for an equivalent dynamic range extension. The required range up to 20\,000\,VEM (Vertical Equivalent Muon, see \cref{sec:calibration}) can be obtained by adjusting the gain in such a way that the ratio of the large to the small PMT signals is 32. 

The SPMT output is required to be linear within 5\% for a peak current up to 50\,mA at a gain of $7{\times}10^5$. All the small photomultipliers have been validated in a test facility by measuring their gain and linearity~\cite{Buscemi}. 
To minimize the number of failures and to ease the maintenance, the SPMT was designed with a passive base, moving the power supply into a separate high voltage power supply (HVPS) module, a custom-made CAEN A7501 HV DC-DC converter, which also provides a measurement of the current flowing through the divider. All the HVPS modules have undergone specific tests to verify their reliability in the challenging environmental conditions and high thermal excursions of the Argentinian Pampa~\cite{Anastasi}.

For consistency with the associated WCD, the dynamic range in the SSD spans from the signal of a single particle, needed for calibration, to large signals, up to ${\sim}2{\times}10^4$\,MIP. The SSD PMT has been chosen accordingly, being linear within 5\% for peak currents up to 160\,mA (for a gain of $8{\times}10^4$).

\subsection{Front-end electronics}

The analog Front-End (FE) has three different configurations, depending on the type of PMT that is connected. For most channels, the amplification of the signal is differential with two amplifier stages. A 7th-order Bessel low-pass filter with 60\,MHz cutoff frequency, designed to preserve the leading-edge timing with minimal distortion of the signal shape, is situated between the two stages.

The signals are digitized by commercial 12-bit 120~MHz dual channel FADCs (Analog Devices AD9628), which achieve this performance with high precision, low noise and minimal power consumption, an important consideration due to the station's small power budget of 10\,W.

The anode channel inputs for the 3 large XP1805 PMTs are split in two and amplified to have a gain ratio of one on the first channel (low gain), and 32 on the second one (high gain).  The anode-channel input for SPMT has a single unitary gain.
The anode channel of the SSD PMT is split in two and amplified to have a gain ratio of 0.25 on the first channel (low gain), and 32 on the second one (high gain). This yields a total gain ratio of 128. The signals are filtered and digitized similarly to the WCD LPMT signals.
The SPMT anode signal is also digitized with 12\,bit at 120\,MHz in a separate channel. The overlap in the dynamic range of LPMT and SPMT is ${\sim}$7\,bit which is sufficient to obtain the cross-calibration for SPMT (see \cref{sec:calibration}).

\begin{figure}
\centering
\includegraphics[width=\textwidth]{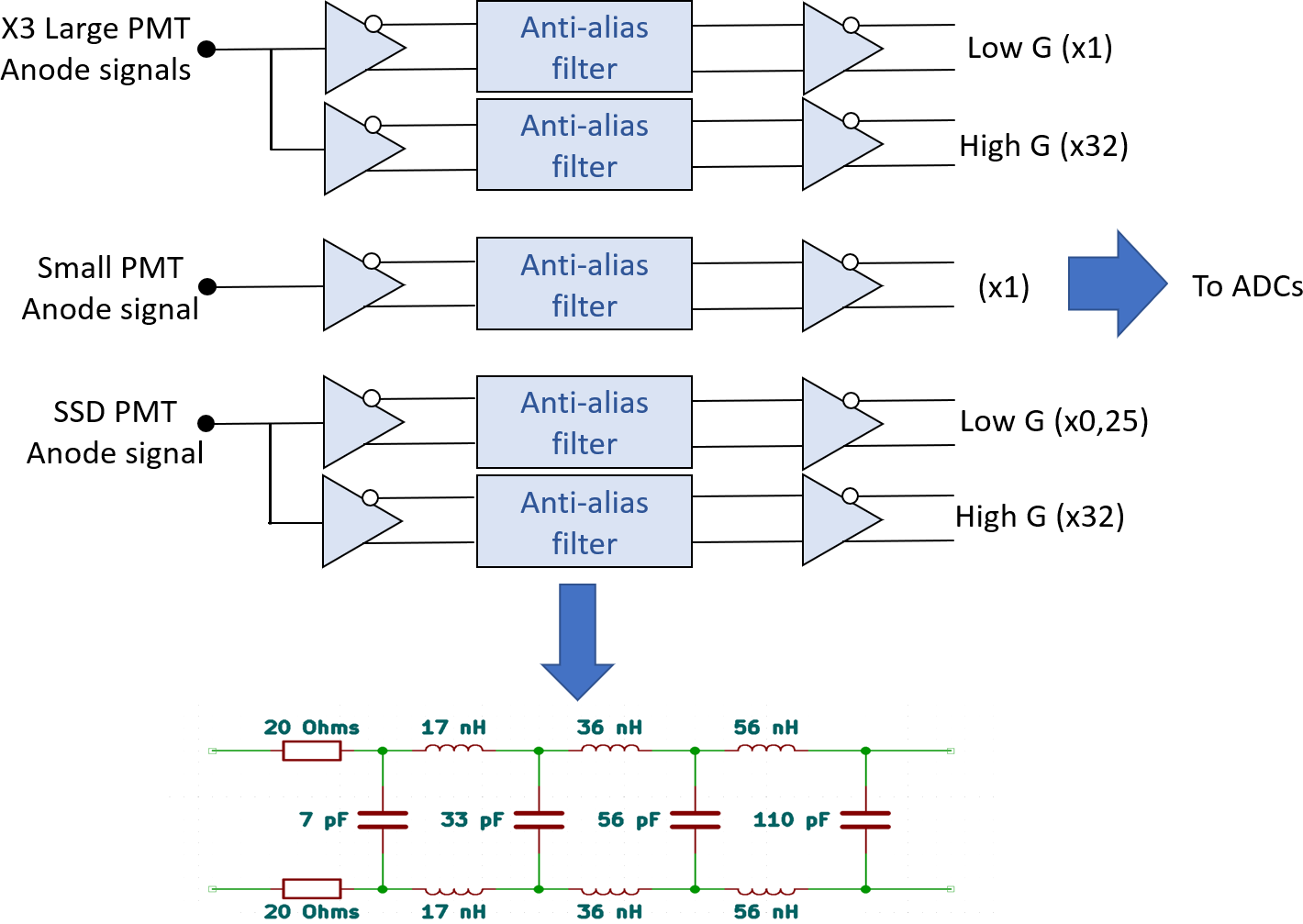}
\caption{Block diagram of the front-end electronics. The total gain factors are indicated.}
\label{fig:FE_block}
\end{figure}

\begin{figure}
\centering
\includegraphics[width=\textwidth]{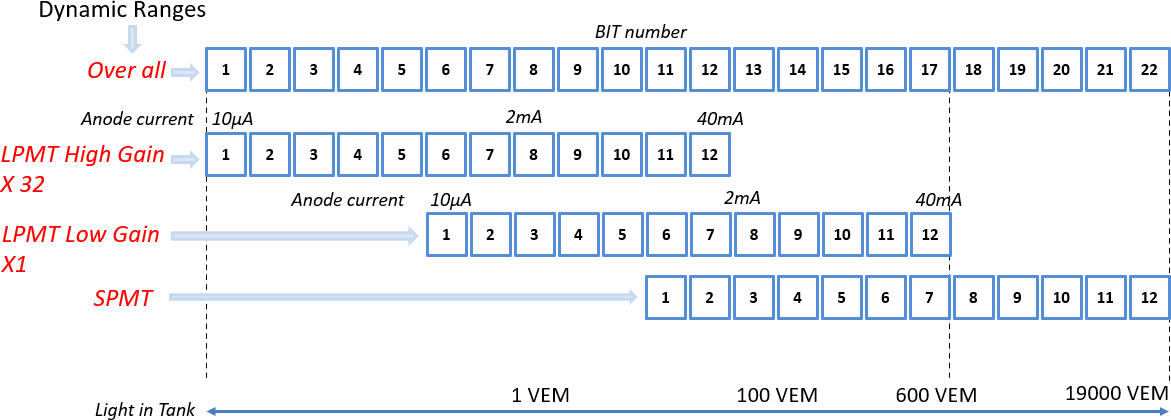}
\caption{Overlap of dynamic ranges.}
\label{fig:Dynamic}
\end{figure}

A block diagram of the front-end electronics channels is shown in \cref{fig:FE_block} and  a scheme of the dynamic ranges is shown in \cref{fig:Dynamic}. 

The intrinsic electronic noise measured in laboratory on the high gain channels is about 2\,LSB and 1/2\,LSB on low gain channels.

\subsection{Timing}

Synchronization of the detectors is provided by tracking variations of the local 120\,MHz clock with respect to the 1\,PPS signal of the Global Positioning System (GPS).  For the upgraded electronics we have selected the Synergy SSR-6TF timing GPS receivers.  This receiver is functionally compatible with the Motorola Oncore UT+ GPS, the one that was used with the former electronics.  The fundamental architecture of the time-tagging firmware module parallels the time-tagging design concept used in the former electronics and is implemented in the UUB board FPGA. The on-board software for initialization of the time-tagging modules, GPS hardware control, and timing data is similar to the former one, with minor modifications needed for the new UUB hardware.  The manufacturer claims an intrinsic GPS device accuracy after the applied granularity correction (the so-called negative saw-tooth) of $\sim$2\,ns.  

The timing performance of the SSR-6FF GPS receiver has been verified in the laboratory, relative to an FS275 GPS-disciplined rubidium atomic clock.  The one-standard-deviation \emph{absolute} timing accuracy is found to range from 2.3\,ns when measured over timescales of a few seconds to about 6\,ns when measured over timescales of several hours.

More importantly, the \emph{relative} timing accuracy (variance on timing of common signal between two SSR-6TF receivers) is measured, to range from better than 1.8\,ns within a temperature-controlled environment to 2.1\,ns when measured in a thermal chamber where temperatures variations are programmed to simulate those expected on the Observatory site (from $-20^\circ$C and up to $+70^\circ$C under the electronics dome).  Additionally, a laboratory test stand that reproduces the time-tagging architecture as implemented in the UUB, was developed.  This test stand is used to verify the timing accuracy and measure any timing offsets for each receiver before it is deployed to the field.  Results from measurements show relative timing accuracy ranging between 4 and 6\,ns.   

The verification of the timing performance of GPS SSR-6TF receivers deployed in the field within UUB prototypes was done by using a synchronization cable to send timing signals between two closely positioned ($\sim$20\,m) UUB-equipped SD stations.  Using this method, a timing accuracy of about 5\,ns was achieved, a result consistent with the lab measurements and the timing granularity as implemented on the UUB.

\subsection{Control and monitoring}

A powerful 16-bit RISC CPU ultra-low-power micro-controller (MSP430) is used for the control and monitoring of the PMT high voltages, the supervision of the various supply voltages and the reset functionality. The power-on sequence of the several supplies for the FPGA is quite complex, and is also controlled by the micro-controller. This device is optimized for low power budget environments.

For all these purposes, it controls 16 logic I/O lines, steers a 12\,bit digital-to-analog converter (DAC) with eight analog outputs, and senses through multiplexers up to 64 analog signals with its internal 12\,bit ADC. The MSP430 also provides a USB  interface, which can be used to monitor and control the various power supplies through a command line interface.  This is used for maintenance. The MSP430 is tied via an I$^2$C-bus to an 256~kbit EEPROM and a pressure/temperature/humidity on board sensor. The system is also in charge of managing the master reset, part of the watchdog and the radio reset. The slow-control is able to restart the UUB after a low battery state due to long bad weather conditions (typically one week without sun on the Observatory site). 

\begin{figure}
\centering
\includegraphics[width=0.70\textwidth]{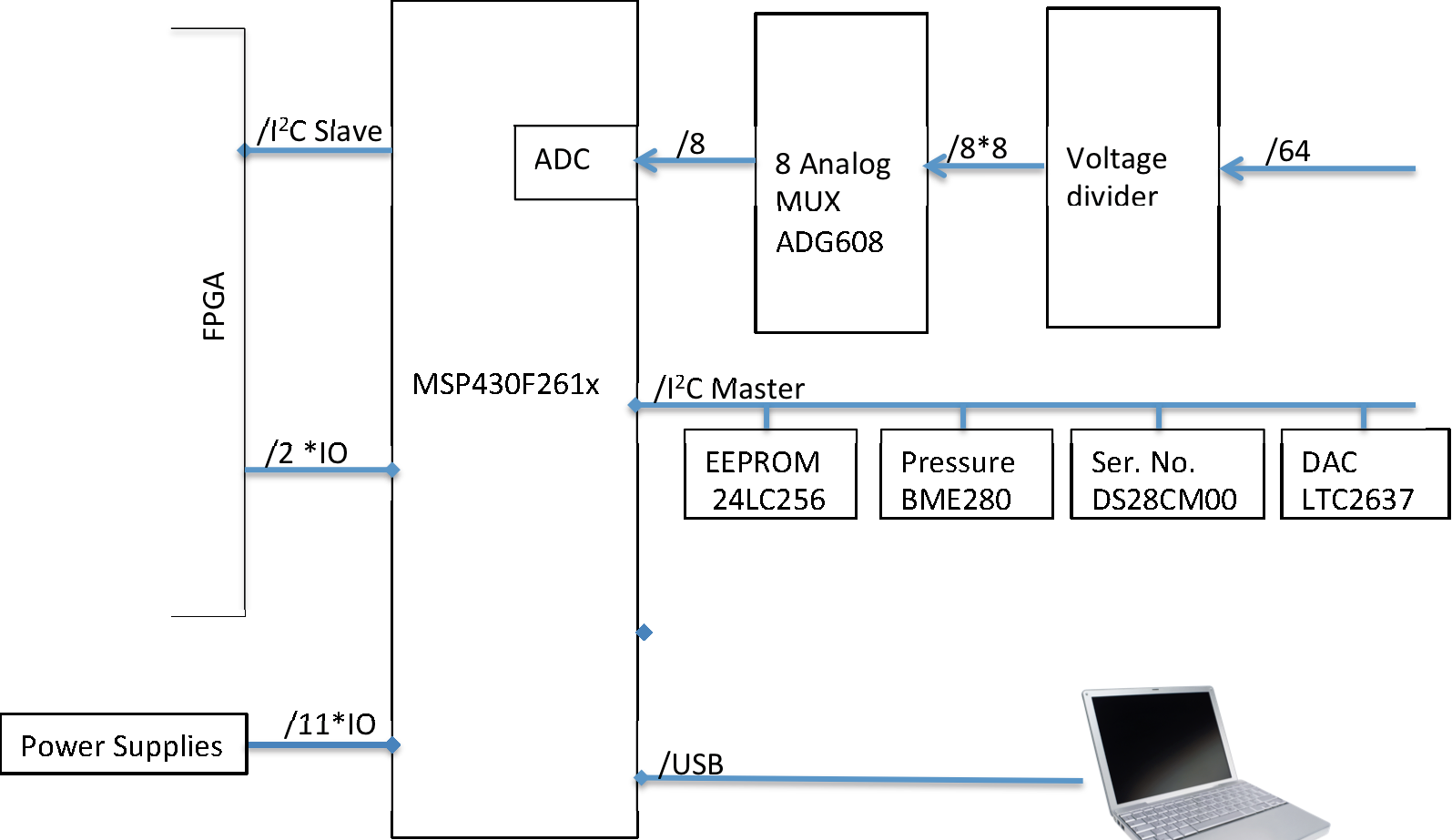}
\caption{Block diagram of the slow-control system.}
\label{fig:SC_block}
\end{figure}

More than 90 monitoring variables, including currents and voltages of the power supplies and the PMTs, are managed by the slow-control firmware and stored in a central database. The firmware also includes diagnostics and safety features. 
The block diagram of the slow-control electronics is shown in \cref{fig:SC_block}. 

\subsection{Firmware and trigger implementation}

The heart of the UUB is a Xilinx Zynq-7020 All Programmable SoC (Artix-7 FPGA and associated Cortex
A9 Dual 333\,MHz ARM co-processor) instead of the older Altera Cyclone series FPGAs used in the previous electronics. Whereas the logic code of the previous FPGAs is written in an Altera specific variant of the  hardware description language VHDL called AHDL, the logic code of AugerPrime version is primarily written in IEEE standard synthesizable
Verilog.  Xilinx Vivado is used for the overall framework, and for standard modules such as memories,
UARTs (Universal Asynchronous Receiver Transmitter), and processor bus interfaces. Xilinx PetaLinux runs on the embedded ARM processor.


The FPGA implements in programmable logic basic digital functions like the readout of the ADCs, the generation of
triggers, and the interfaces to the LED flasher, GPS receiver, and memories. High-level functions like data handling and interactions with the communications radio transceiver are implemented under Linux.  
The addition of
accessible trigger IN/OUT signals and high-speed USB facilitates tests both in the laboratory and on the Observatory site.

A multi-level triggering scheme is used.  The lowest trigger level for each trigger type is denoted T1.  This is formed by the programmable logic and causes the  traces to be transferred to the ARM processor.  The higher level triggers (T2, T3, \ldots) are performed in software and discussed in \cref{sec:localdaq}.

The previous local triggers~\cite{Nitz:1998tk,Nitz:2004be,Abraham:2010zz} (threshold trigger, time-over-threshold trigger (ToT), time-over-threshold deconvolved (ToTd), multiplicity of positive steps (MoPS) trigger) are transferred to new electronics. 
The ToT trigger requires an extended duration signal.   The ToTd variation of the ToT removes to first order the tails of signals from a single particle due to multiple reflections from the station walls.  The MoPS trigger aims to do a similar operation by only looking at the rising edge of signals.  All of these have higher purity and are more efficient for electromagnetic showers and in stations away from the shower core than the simple threshold trigger.
The triggers are implemented by using digitally filtered and down sampled waveforms to reproduce the previous trigger characteristics. 
This consists of taking the full-band traces of UUB with 2048
bins and filtering them, using an FIR Nyquist filter with a 20\,MHz cut-off, to approximate the
frequency response of the previous electronics. In addition, to reproduce the sampling at 40\,MHz of the former electronics, the FADC traces are down-sampled by choosing every third bin on which to apply the trigger algorithm used in the former electronics.
This allows detectors with the new electronics to behave identically to the former configuration at the trigger level and allows deployment of new electronics during the maintenance of the existing system without disturbance to the data taking. To distinguish these down sampled triggers from newer triggers that utilize the full ADC sampling, we include the modifier ``compatibility''. The T2-rates are about 20~Hz for previous and new electronics and the shower trigger (ToT) rates are around one Hz for both electronics.

The increased local processing capabilities allow new
triggers, targeted to neutral primaries, to be implemented such as asymmetry-based triggers, and combined SSD and WCD triggers. Short traces triggered by muon-like signals are stored in so-called ``muon buffers''.  These buffers are read into the processor to facilitate online calibration.  Scalers keep a continuous record of a ``scaler trigger'' rate, and are used to search for correlated increases in rate across the array.  A ``random'' trigger allows acquisition of background data to assist in noise characterization and trigger design.  Finally, the FPGA allows playback of previously recorded or simulated traces to test and verify the implemented trigger algorithms.

\subsection{Local processing software}
\label{sec:localdaq}

\begin{figure}
\centering
\includegraphics[width=0.99\textwidth]{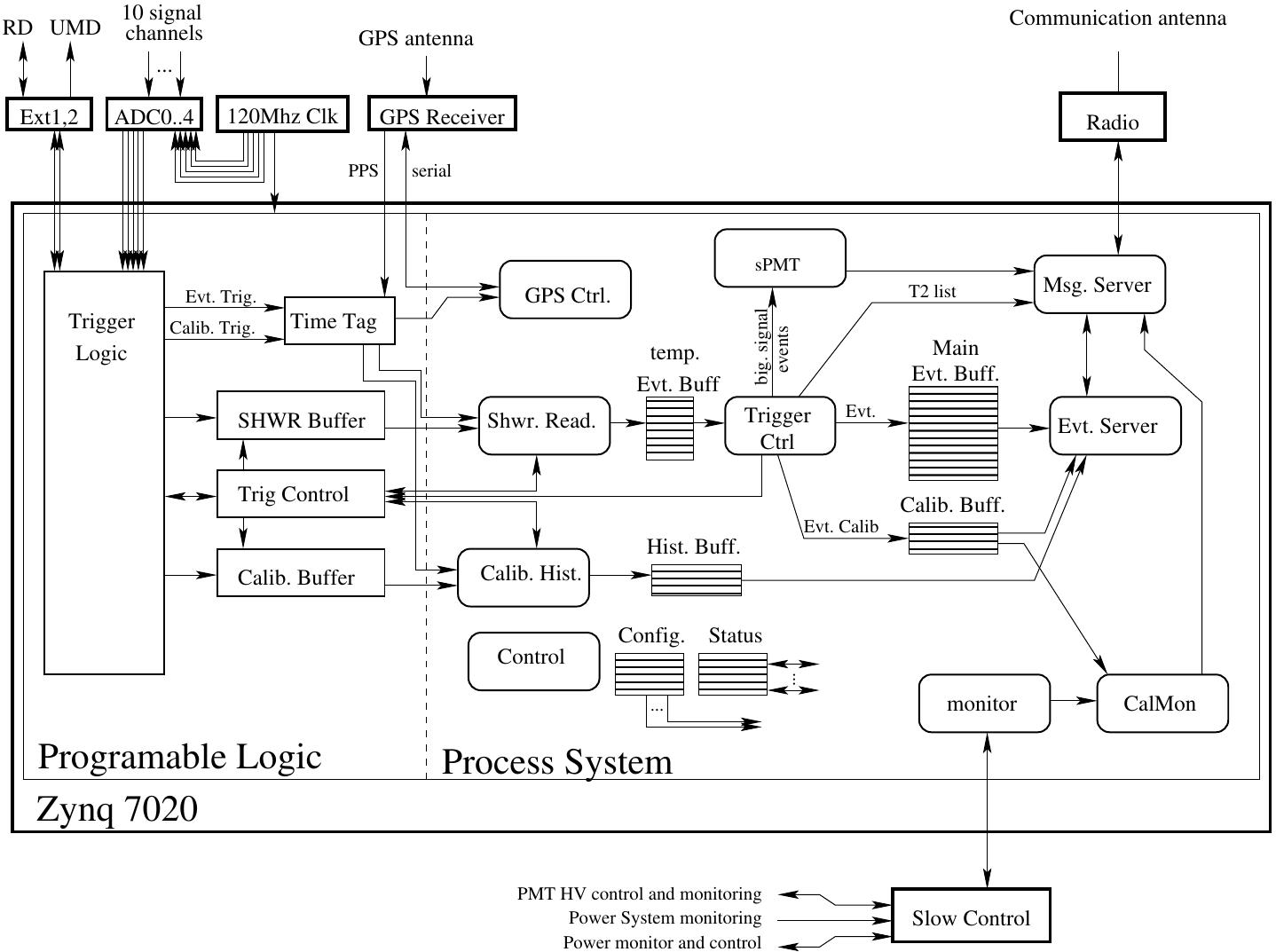}
\caption{Block diagram of local processing Software. The rounded corner blocks are the processes which run on the operating system.}
\label{fig:DAQ_struct_view}
\end{figure}

The speed of the upgraded CPU is more than 10 times faster than that of the previous one, Power PC 403GCX~\cite{NIM}, with a similar increase in memory. This allows more sophisticated processing in the local station. The previous UB code, which used the OS9 operating system, has been ported to Linux.  In this process, the code was adjusted to account for the differences in OS9 and Linux system calls and for the different design structures in the UUB.
\cref{fig:DAQ_struct_view} gives an overview of the local processing software implemented in the UUB. 

In the following, a short description of the local processing software is given. The short names refer to those used in \cref{fig:DAQ_struct_view}.

The data satisfying the T1-trigger condition in ``SHWR Buffer'' is transferred to a temporary event buffer (``temp.\ Evt.\ Buff.'') in the RAM memory. The process ``Trigger Ctrl'' determines if the event passes the T2-trigger condition and calculates the calibration parameters. In case the T2-trigger condition is fulfilled, the event is copied to the main event buffer (``Main Evt.\ Buff.''). 

To decouple trigger rates from station-to-station and PMT-to-PMT gain variations, (most of) the trigger thresholds are computed as a multiple of the most probable peak value of the background vertical equivalent muons ($\text{VEM}_\text{pk}$) generated in each PMT (see \cref{sec:calibration}).  The calculation of $\text{VEM}_\text{pk}$ by the ``Trigger Ctrl'' process proceeds as follows: It starts by setting $\text{VEM}_\text{pk}$ to a default value. With this value the threshold above the baseline for each PMT is set as
\begin{equation}
  \text{Th}_\text{pmt}^\text{type} = \alpha^\text{type} ~ \text{VEM}_\text{pk}
\label{trigger_threshold}
\end{equation}
where $\alpha^\text{type}$ is a constant which depends on the trigger type. For compatibility single-bin trigger threshold,\footnote{The $\text{VEM}_\text{pk}$ for compatibility mode triggers is calculated using filtered and down sampled signals.} the value for $\alpha^\text{T1}=1.75\,\text{VEM}_\text{pk}$.

After this, ``Trigger Ctrl'' determines if the signal passes the T1 condition and calculates
the rates for those PMTs that pass the threshold $\text{Th}_\text{pmt}^{70\,\text{Hz}}$, where $\alpha^\text{{T70}}=2.5$\,VEM (e.g.\ at a threshold of 2.5\,VEM the rate has been found empirically to be 70\,Hz). In case the rate is lower (higher) than 70\,Hz, the $\text{VEM}_\text{pk}$ is decreased (increased) and the thresholds are reset following the \cref{trigger_threshold}.
After some iterations, $\text{VEM}_\text{pk}$ stabilizes to the value that corresponds to the PMT gain.  At this value, the T1 rate is $\sim$100\,Hz and the T2 rate ($\alpha^\text{T2}=3.2$\,VEM) is $\sim$20\,Hz~\cite{Bertou}.

The timestamps of all T2 events (``T2 list'') are sent to the process ``Msg.\ Server'' which at the end  sends the message to Central Data Acquisition System (CDAS). Furthermore, the ``Msg.\ Server'' is responsible to transmit all the messages from all the processes to CDAS ordered by priority, following the radio protocol. It also receives the messages from CDAS and delivers them to the corresponding processes.

When CDAS finds a coincidence between different stations in the ``T2 list'', it emits a level 3 (T3) trigger which goes to the ``Evt.\ Server'' of the corresponding stations. This process gets the event from the ``Main Evt.\ Buff.'', adds the calibration information and histograms, and sends the complete event information back to CDAS.

Short traces which are acquired from the muon buffers by ``Calib.\ Buffer'' are used by ``Calib.\ Hist.''\ to construct histograms of signal amplitude and charge.
The ``CalMon'' process collects the calibration data as well as the power system monitoring data through the process ``monitor'' and reports them periodically to CDAS.

The process ``Control'' searches for the acquisition configuration and stores it  in the ``Config.''\ structure which is shared with all the other processes. Besides this, it oversees the processes through the ``Status'' shared structure. All the acquisition processes update their own information, so that ``Control'' can identify possible problems.


In addition to sending T2 timestamps, event traces, calibration, and monitoring data to CDAS, as well as accepting T3 requests from CDAS, the communication protocol allows sending files and even arbitrary Linux commands from CDAS to selected stations or as a broadcast.  This allows updating the local processing software, and even the compiled programmable logic ``bitstreams'' for the UUB and RD.

\subsection{Implementation and interfaces}

\begin{figure}
\centering
\includegraphics[width=\textwidth]{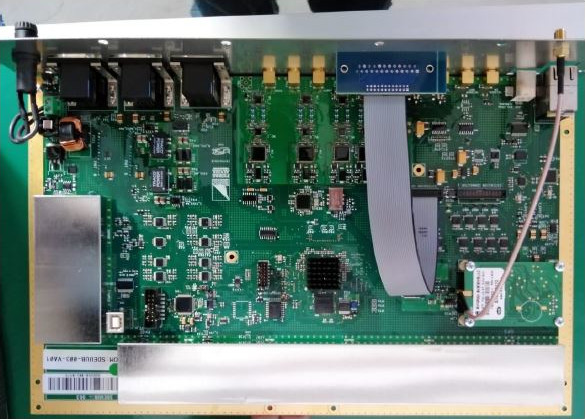}
\caption{Assembled UUB, equipped with front panel, GPS receiver, cables and shielding covers.}
\label{fig:AssembUUB}
\end{figure}

All the functions described above except for the GPS receiver, have been gathered on a single board of $340\,\text{mm}\times215$\,mm size. The printed circuit board (PCB) is a ten copper layer FR4 class 6 board.  The board is fully coated after assembly, on both external sides and edges, using a silicon removable coating product, including  UV marker and RoHS-2 compliant (hazardous substances free, following the European directive). This is done to protect the board against the harsh environment (temperature variations from $-20^\circ$C to $+70^\circ$C under the electronics dome cover, air salinity and humidity).
A photo of the assembled UUB is shown in \cref{fig:AssembUUB}.



The UUB, together with the GPS receiver board, are mounted inside the existing metallic RF-proof enclosure. A new front panel is designed, integrating existing and new connectors for the additional detectors and features. This allows us to keep the current mechanical components of the SD detectors.

\begin{figure}
\centering
\includegraphics[width=0.99\textwidth]{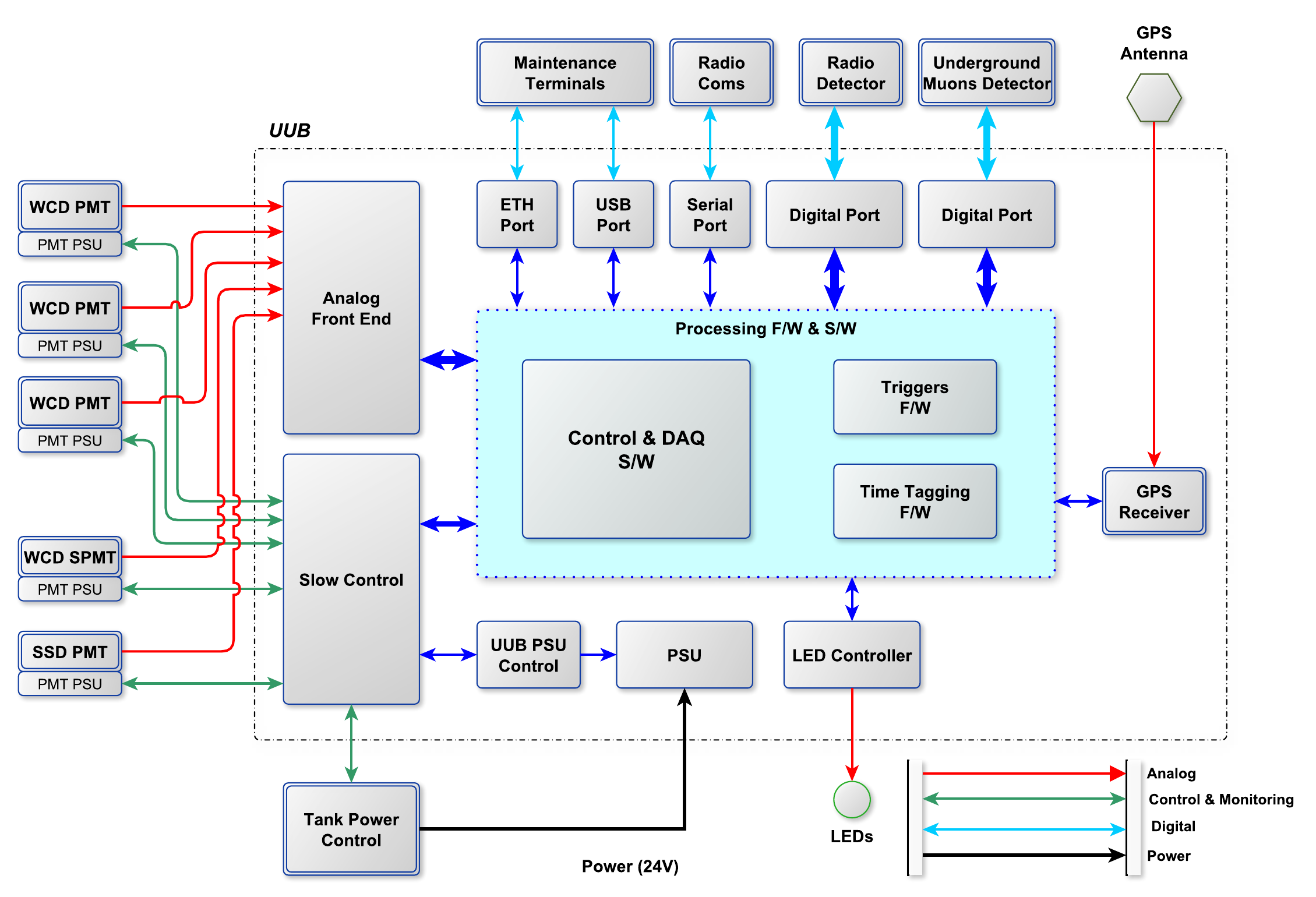}
\caption{Electrical interfaces of the Upgraded Unified Board.}
\label{fig:Interf}
\end{figure}

Two 8\,bit digital ports are provided for additional detectors. The UUB is interfaced with the actual communication system providing 1200\,bps data transmission rate, and with the power system providing 24\,V from the batteries. The previous power budget of 10\,W is increased up to nearly 20\,W by installing new solar panels on each SD station. Additionally, the electrical design of the UUB is made to reduce the conducted and emitted electromagnetic interference to an acceptable level for the RD system by using appropriate filters and shielding material. \cref{fig:Interf} shows all electrical interfaces.

The embedded software of the UUB is interfaced with the existing radio transceiver, using a proprietary communication protocol, the new GPS receiver, using a communication language identical to the the previous receiver, and the new additional detectors, RD and UMD, interfaced via the digital ports.

\section{Production, tests, and installation}

1700 units of SPMTs were procured from Hamamatsu and separate custom-designed HV modules were procured from CAEN. All the modules were tested in Europe prior to shipment to Argentina (see \cref{subsec:dynamicrange}). 
The production and test strategy of the UUBs is described in detail in the following sub-sections. A short description of the deployments strategy of UUBs, SPMTs together with the SSD PMTs is given in the end of the section.

\subsection{Production strategy}

For the mass production of the UUBs, only one manufacturer was selected to reduce the risks of discrepancies that could occur if UUB batches were produced by different companies. The selected manufacturer was the A4F company (Angel for Future, formerly SITAEL), in Italy.

The items requested to the manufacturer were:
\begin{itemize}
\item Procurement of all the components and materials, except those already procured by the Auger Collaboration.
\item Manufacturing or procurement of the printed-circuit boards.
\item Mounting and assembly of the boards according to instructions provided by the Auger Collaboration.
\item Quality control and testing of the boards according to instructions, test plan and test benches provided by the Auger Collaboration (Manufacturing tests).
\item Packing and shipment of the boards with a delivery to the Observatory according to a staged schedule.
\item Warranty on manufacturing and behavior of the boards for a defined period.
\end{itemize}

\subsection{Tests and verification strategy}

\begin{figure}
\centering
\includegraphics[width=\textwidth]{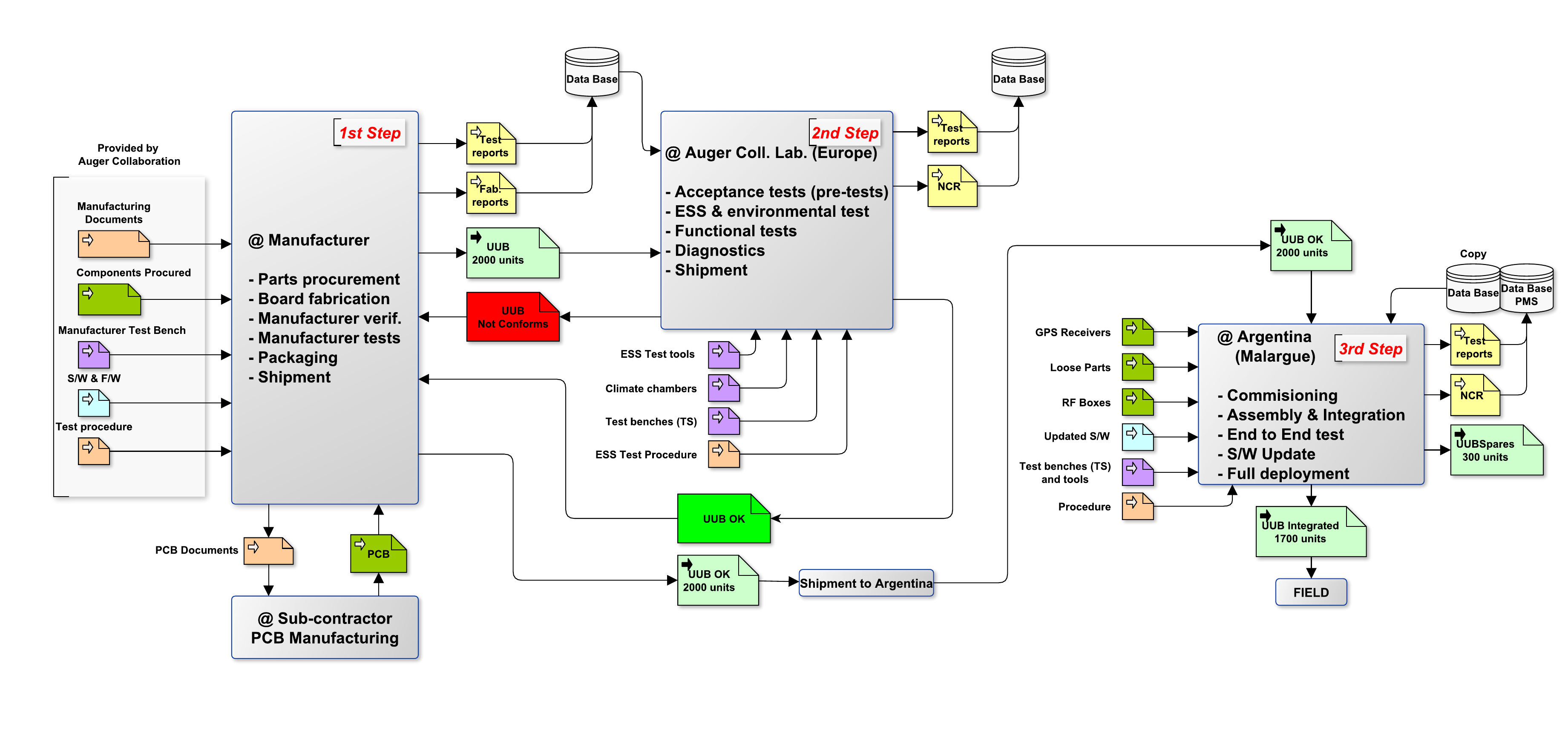}
\caption{Three steps of the UUB testing strategy.}
\label{fig:tests}
\end{figure}

The UUB validation and test process has three steps (see \cref{fig:tests}):
\begin{itemize}
\item The manufacturer test, to verify the proper behavior of almost all the functions of the UUB after assembly, performed at the manufacturer plant.
\item The Environmental and Stress Screening test, to stress the UUB in a climate chamber after manufacturing and to measure performance. This test is performed in Europe, in  a Pierre Auger collaboration laboratory.
\item The final test, performed in Malarg\"ue after delivery, together with the final assembly and before the deployment on site.
\end{itemize}

Therefore, three types of test benches have been developed, each one designed to perform one of the tests described above.

\subsection{Manufacturing tests}

The manufacturer test aims at verifying that all functional blocks of the UUB are correctly assembled and in operation. Two identical test benches were developed and installed at the manufacturer site. Each of them allows to test one UUB at a time.

\begin{figure}
\centering
\includegraphics[width=\textwidth]{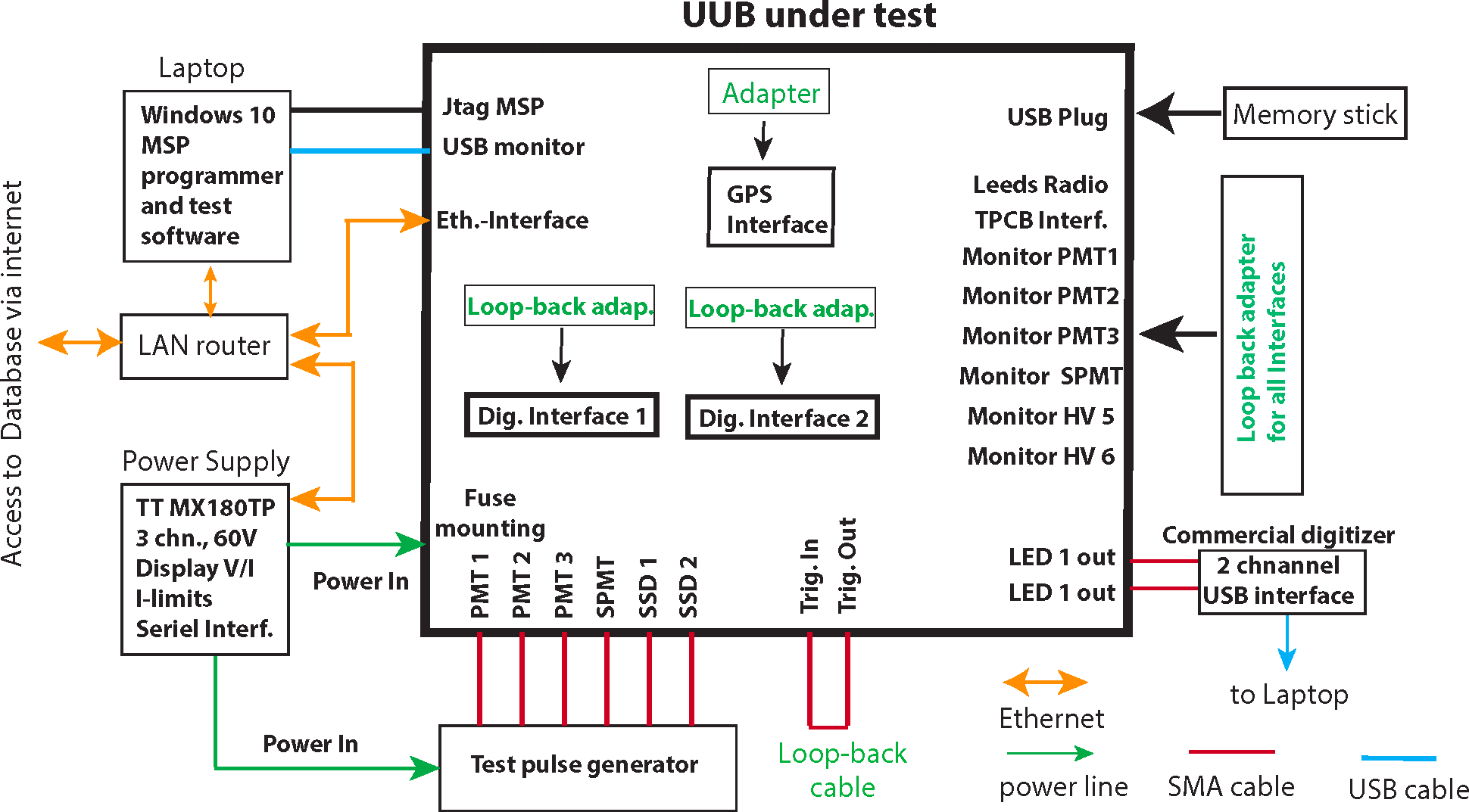}
\caption{Test bench: Loop-back adapters in green, SMA quick fit connectors in red.}
\label{fig:Man}
\end{figure}

The UUB to be tested is mounted on a plastic support frame  and locked with two clamps to a support structure. It is connected to a multi-channel pulse generator through SMA quick-fit connectors mounted on a slider, and it is powered by a programmable power supply. All test results are recorded via an Ethernet interface
or via a digital oscilloscope. Adapters are connected to the UUB during the test to provide specific voltage levels or feedback control signals.
\cref{fig:Man} shows the schematics of the manufacturer test bench.

As a first step of the testing procedure, all UUBs must pass an initial automated optical inspection (AOI) with a system provided by the manufacturer. The inspection can detect problems related to the soldering process (such as excessive or insufficient solder paste) and issues related to component assembly (such as missing components, wrong orientation or distortion of integrated circuits, wrong component polarity) with high efficiency.

Once the automatic procedure is complete, the operator moves the board to a semi-automatic test bench. After connecting all inputs to the test system, a script is executed to install updated firmware for the MSP microcontroller and the FPGA. The UUB then reboots and is ready for the full functionality test. Through a web page it is possible to execute specific tests on the UUB using the Application Programming Interface (API) running under PetaLinux. The test results are automatically analysed and the data is loaded into the web page for inspection. Information about the configuration of the UUB is automatically acquired, formatted and saved together with the test results in a database. 
The database allows the export of results into spreadsheets to produce statistics about parameter variations (e.g.\ voltages and currents). 

\subsection{Environmental stress screening}

After the manufacturing test, the UUBs are submitted to an Environmental Stress Screening (ESS) which is performed to characterize the behaviour of the new electronics under changing environmental conditions typically observed at the Observatory site and to provoke early failures. ESS tests consist of a burn-in procedure followed by temperature cycling, using a dedicated climate chamber. A batch of ten UUBs can be submitted to this test at a time.
 During the burn-in UUBs are subject to rapid temperature changes for 24 hours. Noise, baselines and temperature readings are monitored regularly. This is followed by 10 cycles, from $-20^\circ$C to $+70^\circ$C (temperature change of $3^\circ$C/min). At five temperature points the performance of each UUB is monitored. The tests performed include noise, baseline and linearity dependence on temperature, stability of the ADCs and the anti-alias filter, and over/under voltage protection test.

\begin{figure}
\centering
\includegraphics[width=0.8\textwidth]{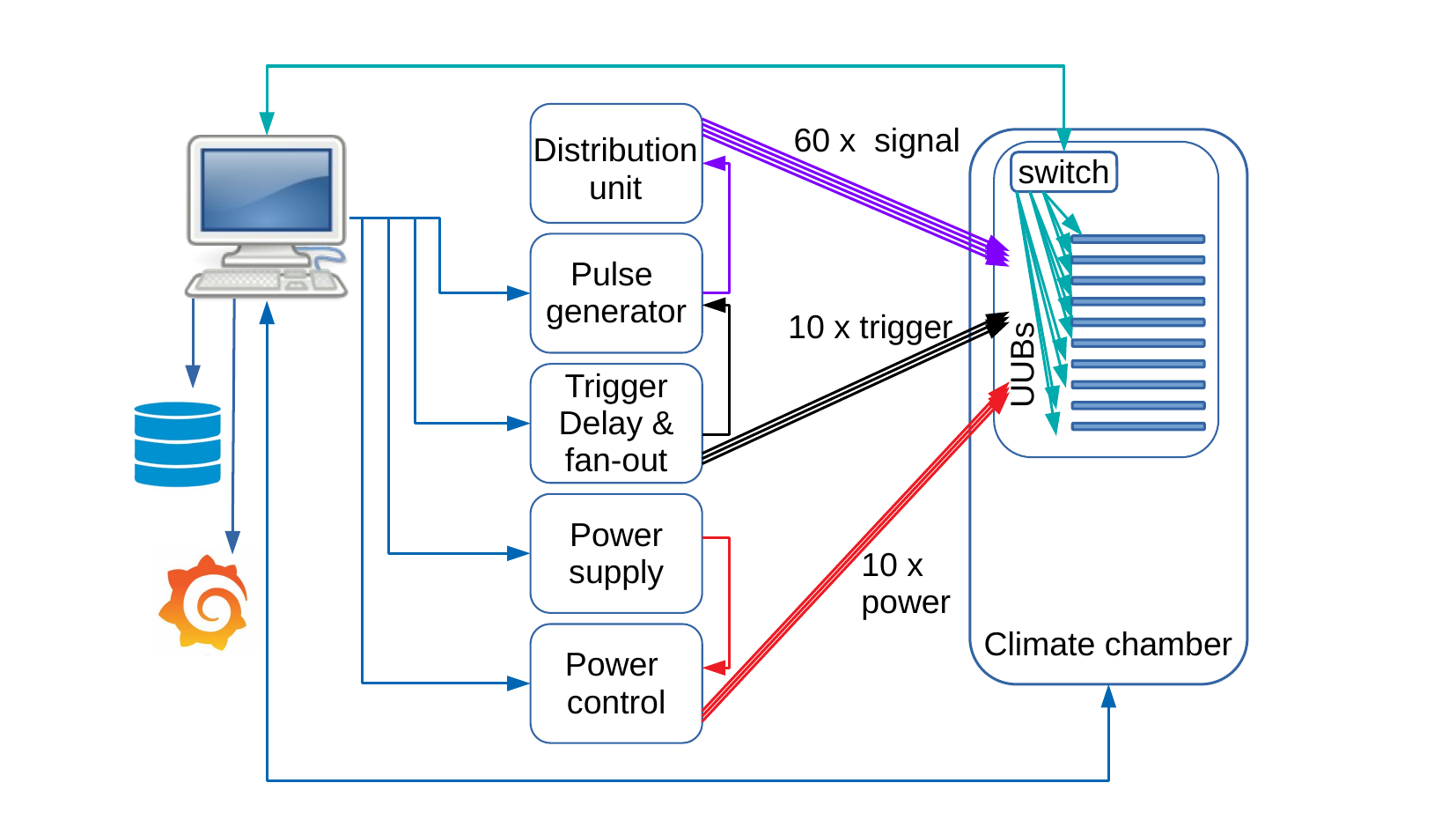}
\caption{ESS test bench scheme. The arrows indicate the information flow: communication with the instruments (USB or ethernet) is depicted in blue, analog test signal is shown in violet, test data acquisition is shown in turquoise. Powering of the devices under test is represented in red.}
\label{fig:ess}
\end{figure}

The scheme of the ESS test bench is shown in \cref{fig:ess}. Communication with the boards is done via Ethernet connection through 
a Gigabit switch placed inside the climate chamber. The test signal is issued by a function generator (AFG3252C, Tektronix), amplified/attenuated and distributed into 60 channels via a custom-made distribution unit. To power the boards, 
a commercial power supply is used, passing through another custom interlock unit allowing monitoring of the current drawn by individual boards and switching the boards off one by one in case of a failure.
The last custom-made unit generates the trigger signal for all ten boards as well as for the pulse generator, with an appropriate time offset.

The handling of the UUBs follows all the ESD-safe precautions specified by the IEC~EN 61340-5-1 standard. Further details on the tests performed can be found in~\cite{Bohacova}.

The complete test procedure is fully automatic, takes 45 hours and is monitored online using the Grafana package (\url{https://grafana.com}), which permits observation from any part of the world. All tests results are summarized in a database.

Among the most frequent failures encountered during the ESS test process, the following issues can be listed:
\begin{itemize}
\item The ADCs were not correctly initialized after rebooting at $70^\circ$C.
This problem is overcome by a software patch, which re-initializes the ADCs when a stuck value is identified.
\item Flipping ADC bits were observed especially at low temperatures. This issue was attributed to faulty components amounting to about 2\% of the batch. Affected ADCs were replaced.
\item Baseline instabilities were occurring mainly at high temperatures. However, this issue should not affect the data since the baseline is taken from the same trace as the signal.
\item Some 3.3\,V DC/DC converter failures were also traced to faulty components, which were replaced.
\item Other, less frequent faults (individual cases) were detected such as soldering issues, broken/missing components or 
booting problems.
\end{itemize}

\subsection{Assembly, final verification, and deployment}

When the UUBs are delivered at the Pierre Auger Observatory site, they undergo a verification process before being deployed.
The first step is to visually inspect each board, searching for minor manufacturing issues or transportation damages. Once the UUBs successfully pass this inspection, they are integrated with the GPS receiver module and the so-called loose parts (i.e.\ cables, connectors, front panel, etc.). The complete setup is mounted inside a sturdy metal RF-enclosure, that acts both as a physical protection and also as an electromagnetic shielding for any radiated RF energy from the UUB (especially from switch mode power supplies) that may interfere with other detectors, in particular the RD.

After the assembly, a final end-to-end verification is performed. This phase performs more than 70 measurements and routines, including the communication via Ethernet and USB ports, the connection with the radio transceiver, the monitoring of the ADC signals, power supplies voltages and currents and the functioning of external connectors. Any non-conformance detected during the visual inspection or tests, initiates a more detailed diagnostic process, allowing us to resolve the issues. This process is performed by expert technicians, fully experienced on the UUB and also on the former electronics (UB). All the assembly, test and diagnostic processes are performed in an electrostatic safe environment, following the usual standards (JEDEC, IEC).

The deployment of the electronics boxes in the field encompasses the integration of the new electronics into the SD array and the data acquisition. This is performed by specific technician teams, fully trained to install the UUB together with the SPMT. Severe weather and site constraints can occur in this phase, challenging the optimization of resources and schedule. The deployment rate per team is roughly between 3 and 4 per day.

\section{Calibration}
\label{sec:calibration}

The calibration of the large WCD PMTs is performed by using atmospheric muons. The Vertical Equivalent Muon signal (VEM, the signal corresponding to a vertical muon crossing the WCD in the center) is the reference unit of the WCD high-gain calibrations and was previously determined on a test tank with an external trigger hodoscope to give on average 95 photoelectrons at the cathode of the XP1805 PMTs~\cite{Bertou}. This corresponds to $\sim$1380 integrated ADC counts above the pedestal after signal digitization on the UUB (see \cref{sec:performances}).  

\begin{figure}
\centering
\def\h{0.375}
\includegraphics[height=\h\textwidth]{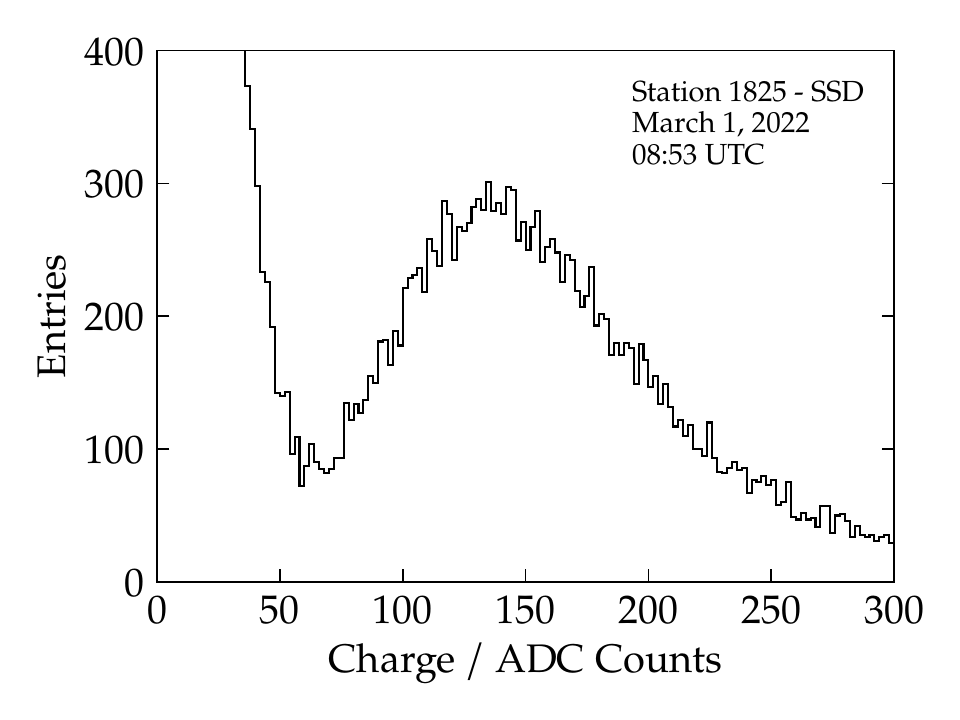}\hfill
\includegraphics[height=\h\textwidth]{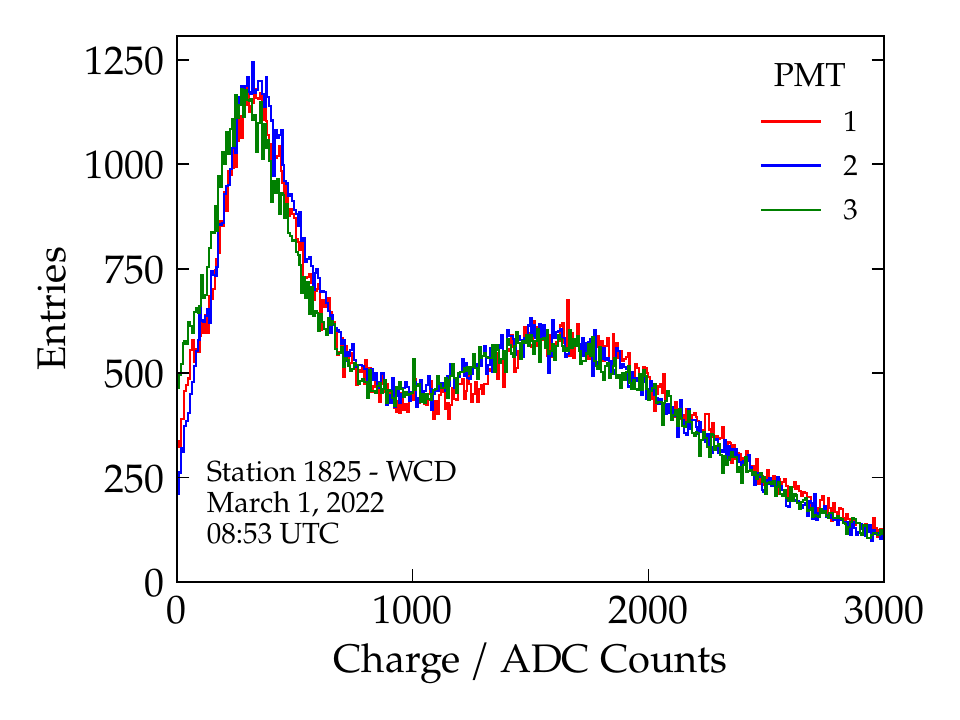}
\caption{Calibration histogram for SSD (left) and the three large PMTs of WCD (right).}
\label{fig:Calib}
\end{figure}

The SSD calibration is based on the signal of a minimum ionizing particle (MIP) going through the detector. About 40\% of the triggered muons of the WCD produce a MIP in the SSD, corresponding to ratio of the SSD surface to the WCD surface. The sensitivity to the muon component used for the calibration can also be increased via a  coincidence calibration between WCD and SSD. 
An example of the MIP and VEM calibration histograms is shown in the left panel of \cref{fig:Calib}. The muon calibration data is continuously recorded and allows us to compensate for the effect of outside temperature variations. The cross-calibration between high gain and low gain channels is set by the electronics design, 32 in the case of WCD and 128 in the case of SSD. This cross-calibration was verified in the ESS test-bench to be $32.2\pm0.3$ for the WCD channels and $126.7\pm1.3$ for the SSD channels (at room temperature and 10\,MHz frequency).

\begin{figure}
\centering
\includegraphics[width=0.8\textwidth]{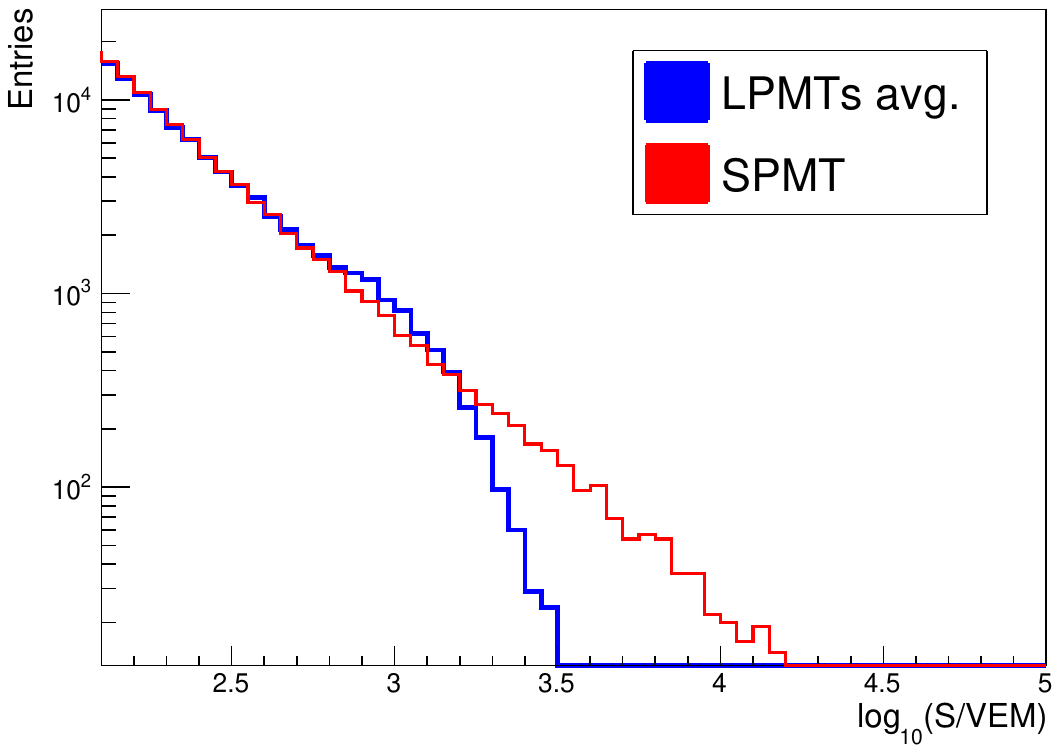}
\caption{Extension of the dynamic range to 20\,000\,VEM using the small PMT. In blue, the average of the signals from the 3 large PMTs. The slight discrepancy between the large and small PMT signals at around 1000 VEM is due to a saturation effect of the large PMT ADCs.}
\label{fig:SPMTCalib}
\end{figure}

Due to its operating parameters, no direct calibration of the SPMT with atmospheric muons is feasible. In this case, the absolute scale in physical units is obtained by cross-calibrating the SPMT using the VEM-calibrated signals of the three large PMTs. A dedicated selection of local small showers\footnote{Small showers are selected requiring a $(n-1)$-fold coincidence among the LPMTs above a threshold between 350 and 550\,FADC counts (depending on the LPMT HV setting), corresponding roughly to 200\,evts/hour.} is setup to this aim, and the cross-calibration is performed in a superposition region limited at the lower end by imposing a minimum threshold of  $\sim$80\,VEM on the WCD PMTs 
to guarantee a reasonably large signal in the SPMT, only marginally affected by statistical fluctuations, and at the higher end by the LPMT saturation. The logarithm of the charge spectrum in one of the upgraded AugerPrime WCD stations is shown in the right panel of \cref{fig:SPMTCalib}. The dynamic range is extended to more than 20\,000\,VEM, as one can see by comparing the unsaturated spectrum from the LPMTs to the one obtained with the SPMT. 
The cross-calibration is performed in 8-hour sliding windows in order to follow the daily evolution of the SPMT gain due to the temperature variations. The choice of 8\,hour intervals assures a precision of about 2.2\%. As such, it can be considered as an optimal trade-off between a large integration period, granting the stability, and a shorter one, needed to compensate for the effects of temperature variations.
The relative difference between the calibrated SPMT and LPMT calibrated signals is always better than 5\% in the whole inter-calibration region. 

The accuracy of the LPMT and SPMT signals contribute to the final SPMT signal accuracy. This final accuracy is better than 5\% above about 3000\,VEM, a value corresponding to the signal produced at 250\,m from the core by showers with energy of $10^{19}$\,eV.

\section{Obtained performances}
\label{sec:performances}

All the SSD detectors have been installed atop the WCDs. The deployment of UUBs, together with SPMTs and SSD PMTs, was completed in July 2023. 
Data taking for commissioning is in progress since the end of 2020. In parallel, the CDAS program, the monitoring program, and the data analysis pipeline have been updated for AugerPrime. The data from AugerPrime is continuously monitored and analyzed to obtain resolutions and to assess the uniformity of detector stations and their long-term performance. In the following, some results obtained in the commissioning studies are presented.

\subsection{Noise performance}

\begin{figure}
\centering
\includegraphics[width=0.5\textwidth]{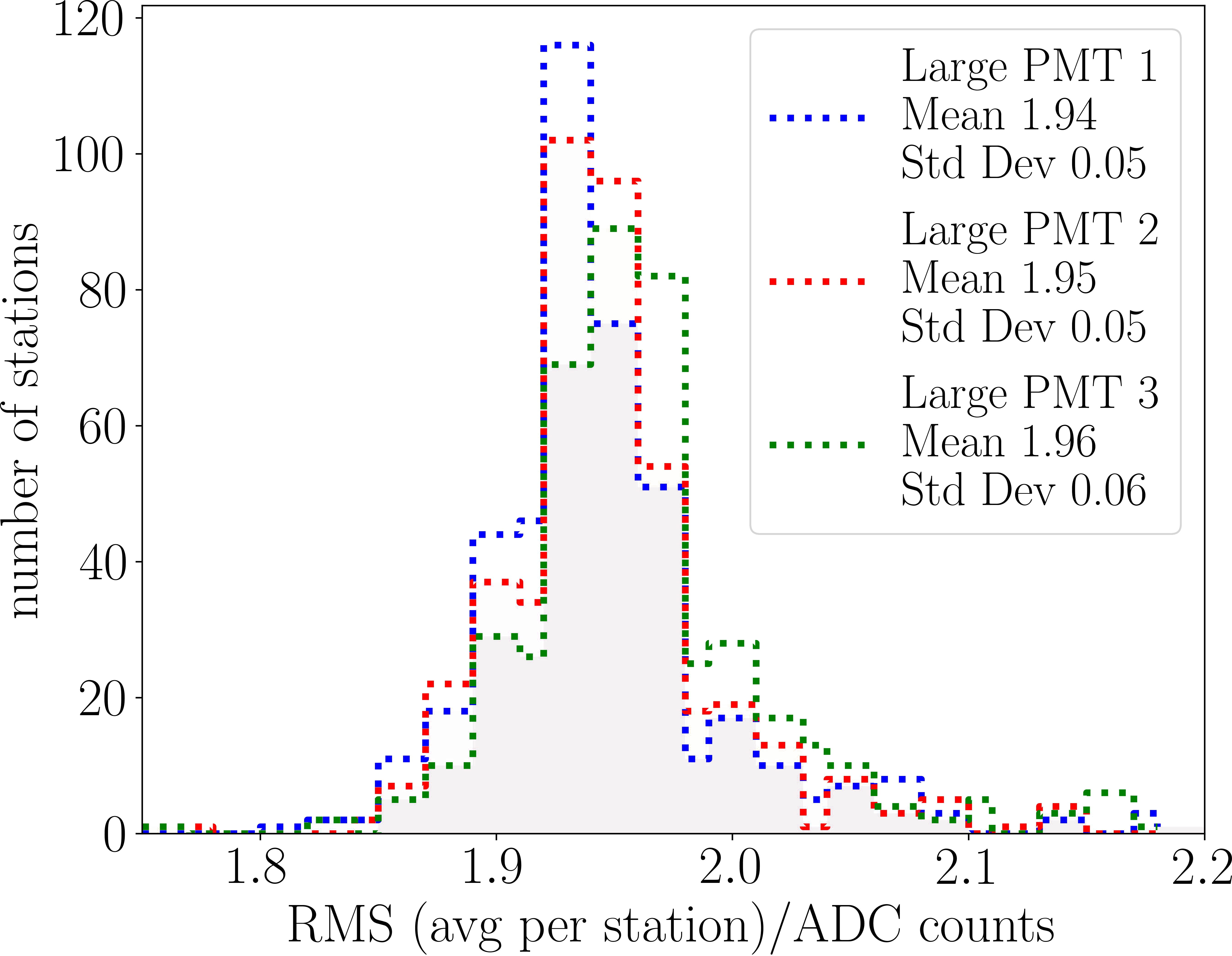}
\caption{The noise of the high gain channel of the three large PMTS.}
\label{fig:LPMT_noise}
\end{figure}

\cref{fig:LPMT_noise} shows the baseline RMS value of the high gain channel of the three large PMTs. The RMS value is an average over about 500 detector stations. 
As can be seen in the figure, the noise for the high-gain channel is below 2\,ADC channels, meeting the requirements. Similarly, the noise of the high-gain channel of the SSD PMTs is below 2\,ADC channels. The SPMT and the low-gain channels of LPMTs and SSD PMTs are well below 1\,ADC channel.

Thunderstorms induce noise in the ADC traces, which increases the trigger rates, and can lead to loss of data if the communications bandwidth becomes saturated. It is currently estimated that this noise would lead to an acceptance loss of about 2\% per year. Studies are in progress to better identify thunderstorm events in order to limit the triggering on noise.

\subsection{Dynamic range}

\begin{figure}
\centering
\includegraphics[width=0.65\textwidth]{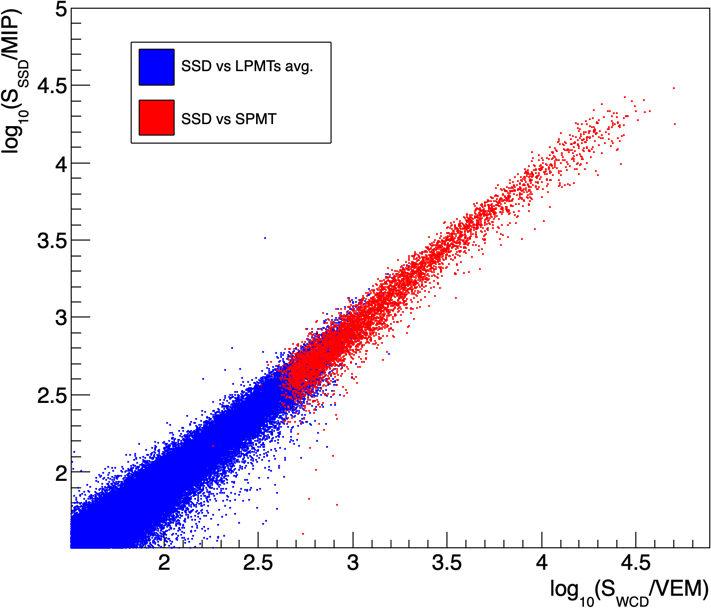}
\caption{Correlation between SSD and WCD signals. The WCD signal are measured up to saturation by the LPMTs (blue dots), and by the SPMT above it (red dots).}
\label{fig:SSD-WCD}
\end{figure}

The excellent correlation between the calibrated signals of the WCD and SSD is shown in \cref{fig:SSD-WCD}, which includes reconstructed data from the Infill region of the SD. Both scales are expressed in the corresponding physics units (VEM for the WCD and MIP for the SSD). The signals in the WCD are measured by the LPMTs up to saturation and by the SPMT in the region above (from $\sim$650 to 20\,000\,VEM and above).
The required dynamic range is reached in both detectors; the obtained correlation clearly shows the validity of the two independent calibrations. 

\subsection{Uniformity and long-term performance}

The various performance parameters are continuously monitored to ensure good detector uniformity and long-term performance. 
The mean charge values measured for VEM and MIP are about 1400 and 110\,ADC channels, respectively. 

The day/night temperature variation can be larger than $20^\circ$C. This induces a typical day/night variations of few  ADC channels for the PMT signals mainly due to the sensitivity of the PMTs to temperature. The muon calibration both for WCD and SSD are made online every minute, allowing the correction for these temperature effects.

\section{Conclusions}

To accommodate new detectors and to increase experimental capabilities, the AugerPrime station electronics has been upgraded. This includes better timing with up-to-date GPS receivers with 5\,ns timing resolution and higher sampling frequency (120\,MHz instead of 40\,MHz) for the ADC traces. Furthermore, a more powerful local processing of the data is obtained by using a Xilinx Zynq-7020 FPGA. The station electronics is gathered on a single board, called UUB. Furthermore, a SPMT is added to WCD detectors to increase the dynamic range. The deployment of the electronics together with the SPMTs was completed mid-2023. 

The test results as well as the commissioning studies show that the design meets the requirements. In particular, the noise for the high-gain channel is below 2\,ADC channels for all PMTs and the results of the commissioning data analysis show good uniformity and stable long-term performance. To reproduce the trigger behavior of the previous electronics, a compatibility mode was designed for UUB triggering in the FPGA firmware. This allows a smooth transition from the previous SD array to the AugerPrime array.

\appendix

\section{Diagram of the UUB architecture}

\begin{figure}
\centering
\includegraphics[height=0.9\textwidth,angle=-90]{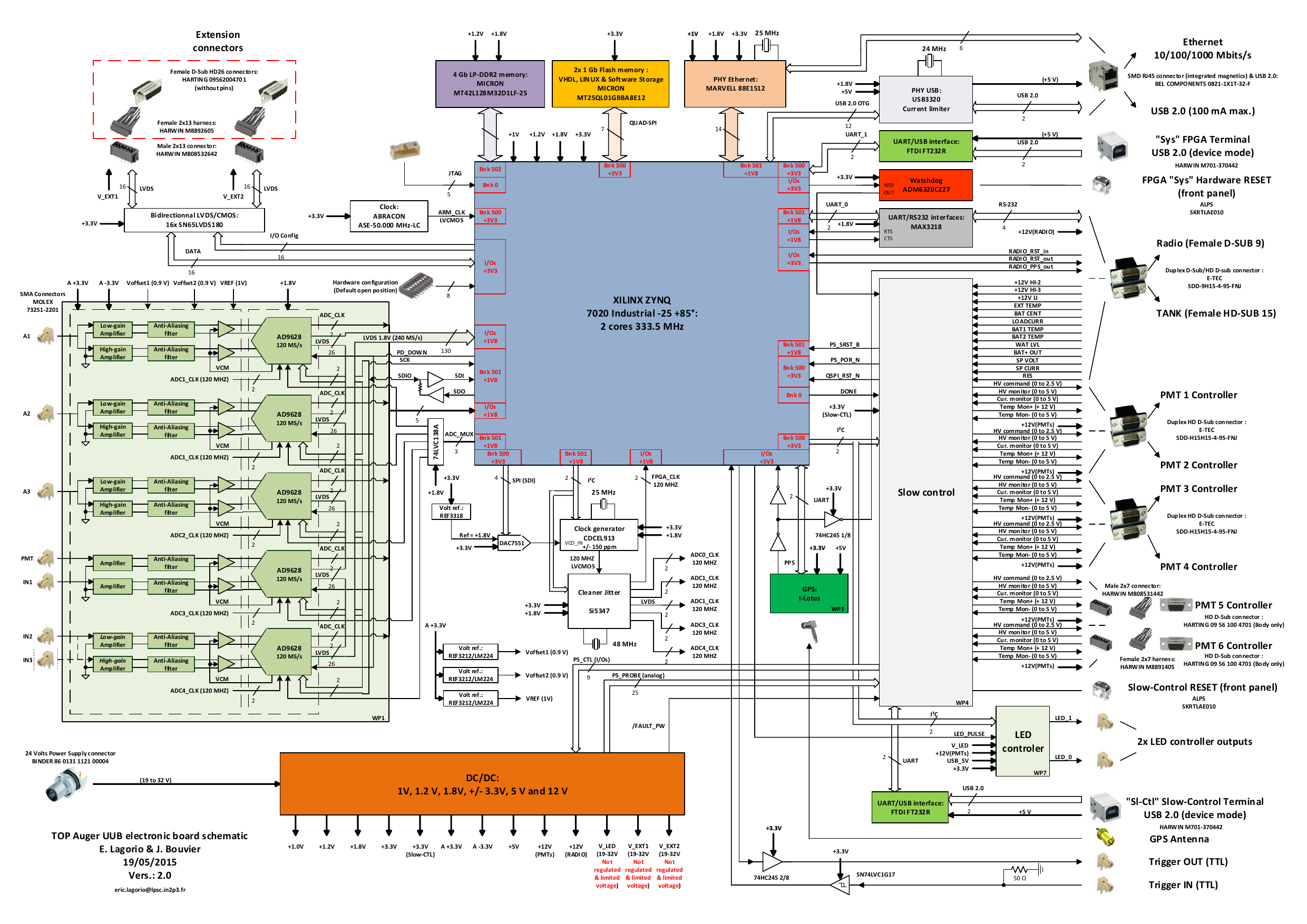}
\caption{Functional diagram of the Upgraded Unified Board.}
\label{fig:Zynq2}
\end{figure}

The \cref{fig:Zynq2} shows a more detailed diagram of the UUB architecture.

\clearpage

\section*{The Pierre Auger Collaboration}
\small

\begin{sloppypar}\noindent
A.~Abdul Halim$^{13}$,
P.~Abreu$^{72}$,
M.~Aglietta$^{54,52}$,
P.~Allison$^{\ell}$,
I.~Allekotte$^{1}$,
K.~Almeida Cheminant$^{70}$,
A.~Almela$^{7,12}$,
R.~Aloisio$^{45,46}$,
J.~Alvarez-Mu\~niz$^{79}$,
J.~Ammerman Yebra$^{79}$
% ...
\end{sloppypar}

\begin{center}
\rule{0.1\columnwidth}{0.5pt}
\raisebox{-0.4ex}{\scriptsize$\bullet$}
\rule{0.1\columnwidth}{0.5pt}
\end{center}

\vspace{-1ex}
\footnotesize
\begin{description}[labelsep=0.2em,align=right,labelwidth=0.7em,labelindent=0em,leftmargin=2em,noitemsep]
\item[$^{1}$] Centro At\'omico Bariloche and Instituto Balseiro (CNEA-UNCuyo-CONICET), San Carlos de Bariloche, Argentina
\item[$^{2}$] Departamento de F\'\i{}sica and Departamento de Ciencias de la Atm\'osfera y los Oc\'eanos, FCEyN, Universidad de Buenos Aires and CONICET, Buenos Aires, Argentina
\item[$^{3}$] IFLP, Universidad Nacional de La Plata and CONICET, La Plata, Argentina
\item[$^{4}$] Instituto de Astronom\'\i{}a y F\'\i{}sica del Espacio (IAFE, CONICET-UBA), Buenos Aires, Argentina
\item[$^{5}$] Instituto de F\'\i{}sica de Rosario (IFIR) -- CONICET/U.N.R.\ and Facultad de Ciencias Bioqu\'\i{}micas y Farmac\'euticas U.N.R., Rosario, Argentina
\item[$^{6}$] Instituto de Tecnolog\'\i{}as en Detecci\'on y Astropart\'\i{}culas (CNEA, CONICET, UNSAM), and Universidad Tecnol\'ogica Nacional -- Facultad Regional Mendoza (CONICET/CNEA), Mendoza, Argentina
\item[$^{7}$] Instituto de Tecnolog\'\i{}as en Detecci\'on y Astropart\'\i{}culas (CNEA, CONICET, UNSAM), Buenos Aires, Argentina
\item[$^{8}$] International Center of Advanced Studies and Instituto de Ciencias F\'\i{}sicas, ECyT-UNSAM and CONICET, Campus Miguelete -- San Mart\'\i{}n, Buenos Aires, Argentina
\item[$^{9}$] Laboratorio Atm\'osfera -- Departamento de Investigaciones en L\'aseres y sus Aplicaciones -- UNIDEF (CITEDEF-CONICET), Argentina
\item[$^{10}$] Observatorio Pierre Auger, Malarg\"ue, Argentina
\item[$^{11}$] Observatorio Pierre Auger and Comisi\'on Nacional de Energ\'\i{}a At\'omica, Malarg\"ue, Argentina
\item[$^{12}$] Universidad Tecnol\'ogica Nacional -- Facultad Regional Buenos Aires, Buenos Aires, Argentina
\item[$^{13}$] University of Adelaide, Adelaide, S.A., Australia
\item[$^{14}$] Universit\'e Libre de Bruxelles (ULB), Brussels, Belgium
\item[$^{15}$] Vrije Universiteit Brussels, Brussels, Belgium
\item[$^{16}$] Centro Federal de Educa\c{c}\~ao Tecnol\'ogica Celso Suckow da Fonseca, Petropolis, Brazil
\item[$^{17}$] Instituto Federal de Educa\c{c}\~ao, Ci\^encia e Tecnologia do Rio de Janeiro (IFRJ), Brazil
\item[$^{18}$] Universidade de S\~ao Paulo, Escola de Engenharia de Lorena, Lorena, SP, Brazil
\item[$^{19}$] Universidade de S\~ao Paulo, Instituto de F\'\i{}sica de S\~ao Carlos, S\~ao Carlos, SP, Brazil
\item[$^{20}$] Universidade de S\~ao Paulo, Instituto de F\'\i{}sica, S\~ao Paulo, SP, Brazil
\item[$^{21}$] Universidade Estadual de Campinas, IFGW, Campinas, SP, Brazil
\item[$^{22}$] Universidade Estadual de Feira de Santana, Feira de Santana, Brazil
\item[$^{23}$] Universidade Federal de Campina Grande, Centro de Ciencias e Tecnologia, Campina Grande, Brazil
\item[$^{24}$] Universidade Federal do ABC, Santo Andr\'e, SP, Brazil
\item[$^{25}$] Universidade Federal do Paran\'a, Setor Palotina, Palotina, Brazil
\item[$^{26}$] Universidade Federal do Rio de Janeiro, Instituto de F\'\i{}sica, Rio de Janeiro, RJ, Brazil
\item[$^{27}$] Universidade Federal do Rio de Janeiro (UFRJ), Observat\'orio do Valongo, Rio de Janeiro, RJ, Brazil
\item[$^{28}$] Universidade Federal Fluminense, EEIMVR, Volta Redonda, RJ, Brazil
\item[$^{29}$] Universidad de Medell\'\i{}n, Medell\'\i{}n, Colombia
\item[$^{30}$] Universidad Industrial de Santander, Bucaramanga, Colombia
\item[$^{31}$] Charles University, Faculty of Mathematics and Physics, Institute of Particle and Nuclear Physics, Prague, Czech Republic
\item[$^{32}$] Institute of Physics of the Czech Academy of Sciences, Prague, Czech Republic
\item[$^{33}$] Palacky University, Olomouc, Czech Republic
\item[$^{34}$] CNRS/IN2P3, IJCLab, Universit\'e Paris-Saclay, Orsay, France
\item[$^{35}$] Laboratoire de Physique Nucl\'eaire et de Hautes Energies (LPNHE), Sorbonne Universit\'e, Universit\'e de Paris, CNRS-IN2P3, Paris, France
\item[$^{36}$] Univ.\ Grenoble Alpes, CNRS, Grenoble Institute of Engineering Univ.\ Grenoble Alpes, LPSC-IN2P3, 38000 Grenoble, France
\item[$^{37}$] Universit\'e Paris-Saclay, CNRS/IN2P3, IJCLab, Orsay, France
\item[$^{38}$] Bergische Universit\"at Wuppertal, Department of Physics, Wuppertal, Germany
\item[$^{39}$] Karlsruhe Institute of Technology (KIT), Institute for Experimental Particle Physics, Karlsruhe, Germany
\item[$^{40}$] Karlsruhe Institute of Technology (KIT), Institut f\"ur Prozessdatenverarbeitung und Elektronik, Karlsruhe, Germany
\item[$^{41}$] Karlsruhe Institute of Technology (KIT), Institute for Astroparticle Physics, Karlsruhe, Germany
\item[$^{42}$] RWTH Aachen University, III.\ Physikalisches Institut A, Aachen, Germany
\item[$^{43}$] Universit\"at Hamburg, II.\ Institut f\"ur Theoretische Physik, Hamburg, Germany
\item[$^{44}$] Universit\"at Siegen, Department Physik -- Experimentelle Teilchenphysik, Siegen, Germany
\item[$^{45}$] Gran Sasso Science Institute, L'Aquila, Italy
\item[$^{46}$] INFN Laboratori Nazionali del Gran Sasso, Assergi (L'Aquila), Italy
\item[$^{47}$] INFN, Sezione di Catania, Catania, Italy
\item[$^{48}$] INFN, Sezione di Lecce, Lecce, Italy
\item[$^{49}$] INFN, Sezione di Milano, Milano, Italy
\item[$^{50}$] INFN, Sezione di Napoli, Napoli, Italy
\item[$^{51}$] INFN, Sezione di Roma ``Tor Vergata'', Roma, Italy
\item[$^{52}$] INFN, Sezione di Torino, Torino, Italy
\item[$^{53}$] Istituto di Astrofisica Spaziale e Fisica Cosmica di Palermo (INAF), Palermo, Italy
\item[$^{54}$] Osservatorio Astrofisico di Torino (INAF), Torino, Italy
\item[$^{55}$] Politecnico di Milano, Dipartimento di Scienze e Tecnologie Aerospaziali , Milano, Italy
\item[$^{56}$] Universit\`a del Salento, Dipartimento di Matematica e Fisica ``E.\ De Giorgi'', Lecce, Italy
\item[$^{57}$] Universit\`a dell'Aquila, Dipartimento di Scienze Fisiche e Chimiche, L'Aquila, Italy
\item[$^{58}$] Universit\`a di Catania, Dipartimento di Fisica e Astronomia ``Ettore Majorana``, Catania, Italy
\item[$^{59}$] Universit\`a di Milano, Dipartimento di Fisica, Milano, Italy
\item[$^{60}$] Universit\`a di Napoli ``Federico II'', Dipartimento di Fisica ``Ettore Pancini'', Napoli, Italy
\item[$^{61}$] Universit\`a di Palermo, Dipartimento di Fisica e Chimica ''E.\ Segr\`e'', Palermo, Italy
\item[$^{62}$] Universit\`a di Roma ``Tor Vergata'', Dipartimento di Fisica, Roma, Italy
\item[$^{63}$] Universit\`a Torino, Dipartimento di Fisica, Torino, Italy
\item[$^{64}$] Benem\'erita Universidad Aut\'onoma de Puebla, Puebla, M\'exico
\item[$^{65}$] Unidad Profesional Interdisciplinaria en Ingenier\'\i{}a y Tecnolog\'\i{}as Avanzadas del Instituto Polit\'ecnico Nacional (UPIITA-IPN), M\'exico, D.F., M\'exico
\item[$^{66}$] Universidad Aut\'onoma de Chiapas, Tuxtla Guti\'errez, Chiapas, M\'exico
\item[$^{67}$] Universidad Michoacana de San Nicol\'as de Hidalgo, Morelia, Michoac\'an, M\'exico
\item[$^{68}$] Universidad Nacional Aut\'onoma de M\'exico, M\'exico, D.F., M\'exico
\item[$^{69}$] Universidad Nacional de San Agustin de Arequipa, Facultad de Ciencias Naturales y Formales, Arequipa, Peru
\item[$^{70}$] Institute of Nuclear Physics PAN, Krakow, Poland
\item[$^{71}$] University of \L{}\'od\'z, Faculty of High-Energy Astrophysics,\L{}\'od\'z, Poland
\item[$^{72}$] Laborat\'orio de Instrumenta\c{c}\~ao e F\'\i{}sica Experimental de Part\'\i{}culas -- LIP and Instituto Superior T\'ecnico -- IST, Universidade de Lisboa -- UL, Lisboa, Portugal
\item[$^{73}$] ``Horia Hulubei'' National Institute for Physics and Nuclear Engineering, Bucharest-Magurele, Romania
\item[$^{74}$] Institute of Space Science, Bucharest-Magurele, Romania
\item[$^{75}$] University Politehnica of Bucharest, Bucharest, Romania
\item[$^{76}$] Center for Astrophysics and Cosmology (CAC), University of Nova Gorica, Nova Gorica, Slovenia
\item[$^{77}$] Experimental Particle Physics Department, J.\ Stefan Institute, Ljubljana, Slovenia
\item[$^{78}$] Universidad de Granada and C.A.F.P.E., Granada, Spain
\item[$^{79}$] Instituto Galego de F\'\i{}sica de Altas Enerx\'\i{}as (IGFAE), Universidade de Santiago de Compostela, Santiago de Compostela, Spain
\item[$^{80}$] IMAPP, Radboud University Nijmegen, Nijmegen, The Netherlands
\item[$^{81}$] Nationaal Instituut voor Kernfysica en Hoge Energie Fysica (NIKHEF), Science Park, Amsterdam, The Netherlands
\item[$^{82}$] Stichting Astronomisch Onderzoek in Nederland (ASTRON), Dwingeloo, The Netherlands
\item[$^{83}$] Universiteit van Amsterdam, Faculty of Science, Amsterdam, The Netherlands
\item[$^{84}$] Case Western Reserve University, Cleveland, OH, USA
\item[$^{85}$] Colorado School of Mines, Golden, CO, USA
\item[$^{86}$] Department of Physics and Astronomy, Lehman College, City University of New York, Bronx, NY, USA
\item[$^{87}$] Michigan Technological University, Houghton, MI, USA
\item[$^{88}$] New York University, New York, NY, USA
\item[$^{89}$] University of Chicago, Enrico Fermi Institute, Chicago, IL, USA
\item[$^{90}$] University of Delaware, Department of Physics and Astronomy, Bartol Research Institute, Newark, DE, USA
\item[$^{91}$] University of Wisconsin-Madison, Department of Physics and WIPAC, Madison, WI, USA
\item[] -----
\item[$^{a}$] Louisiana State University, Baton Rouge, LA, USA
\item[$^{b}$] also at University of Bucharest, Physics Department, Bucharest, Romania
\item[$^{c}$] School of Physics and Astronomy, University of Leeds, Leeds, United Kingdom
\item[$^{d}$] now at Agenzia Spaziale Italiana (ASI).\ Via del Politecnico 00133, Roma, Italy
\item[$^{e}$] Fermi National Accelerator Laboratory, Fermilab, Batavia, IL, USA
\item[$^{f}$] now at Graduate School of Science, Osaka Metropolitan University, Osaka, Japan
\item[$^{g}$] now at ECAP, Erlangen, Germany
\item[$^{h}$] Max-Planck-Institut f\"ur Radioastronomie, Bonn, Germany
\item[$^{i}$] also at Kapteyn Institute, University of Groningen, Groningen, The Netherlands
\item[$^{j}$] Colorado State University, Fort Collins, CO, USA
\item[$^{k}$] Pennsylvania State University, University Park, PA, USA
\item[$^{\ell}$] Ohio State University, Department of Physics, Columbus, OH, USA
\end{description}

\vspace{-1ex}
\footnotesize
\section*{Acknowledgments}

\begin{sloppypar}
The successful installation, commissioning, and operation of the Pierre
Auger Observatory would not have been possible without the strong
commitment and effort from the technical and administrative staff in
Malarg\"ue. We are very grateful to the following agencies and
organizations for financial support:
\end{sloppypar}

\begin{sloppypar}
Argentina -- Comisi\'on Nacional de Energ\'\i{}a At\'omica; Agencia Nacional de
Promoci\'on Cient\'\i{}fica y Tecnol\'ogica (ANPCyT); Consejo Nacional de
Investigaciones Cient\'\i{}ficas y T\'ecnicas (CONICET); Gobierno de la
Provincia de Mendoza; Municipalidad de Malarg\"ue; NDM Holdings and Valle
Las Le\~nas; in gratitude for their continuing cooperation over land
access; Australia -- the Australian Research Council; Belgium -- Fonds
de la Recherche Scientifique (FNRS); Research Foundation Flanders (FWO),
Marie Curie Action of the European Union Grant No.~101107047; Brazil --
Conselho Nacional de Desenvolvimento Cient\'\i{}fico e Tecnol\'ogico (CNPq);
Financiadora de Estudos e Projetos (FINEP); Funda\c{c}\~ao de Amparo \`a
Pesquisa do Estado de Rio de Janeiro (FAPERJ); S\~ao Paulo Research
Foundation (FAPESP) Grants No.~2019/10151-2, No.~2010/07359-6 and
No.~1999/05404-3; Minist\'erio da Ci\^encia, Tecnologia, Inova\c{c}\~oes e
Comunica\c{c}\~oes (MCTIC); Czech Republic -- Grant No.~MSMT CR LTT18004,
LM2015038, LM2018102, CZ.02.1.01/0.0/0.0/16{\textunderscore}013/0001402,
CZ.02.1.01/0.0/0.0/18{\textunderscore}046/0016010 and
CZ.02.1.01/0.0/0.0/17{\textunderscore}049/0008422; France -- Centre de Calcul
IN2P3/CNRS; Centre National de la Recherche Scientifique (CNRS); Conseil
R\'egional Ile-de-France; D\'epartement Physique Nucl\'eaire et Corpusculaire
(PNC-IN2P3/CNRS); D\'epartement Sciences de l'Univers (SDU-INSU/CNRS);
Institut Lagrange de Paris (ILP) Grant No.~LABEX ANR-10-LABX-63 within
the Investissements d'Avenir Programme Grant No.~ANR-11-IDEX-0004-02;
Germany -- Bundesministerium f\"ur Bildung und Forschung (BMBF); Deutsche
Forschungsgemeinschaft (DFG); Finanzministerium Baden-W\"urttemberg;
Helmholtz Alliance for Astroparticle Physics (HAP);
Helmholtz-Gemeinschaft Deutscher Forschungszentren (HGF); Ministerium
f\"ur Kultur und Wissenschaft des Landes Nordrhein-Westfalen; Ministerium
f\"ur Wissenschaft, Forschung und Kunst des Landes Baden-W\"urttemberg;
Italy -- Istituto Nazionale di Fisica Nucleare (INFN); Istituto
Nazionale di Astrofisica (INAF); Ministero dell'Universit\`a e della
Ricerca (MUR); CETEMPS Center of Excellence; Ministero degli Affari
Esteri (MAE), ICSC Centro Nazionale di Ricerca in High Performance
Computing, Big Data and Quantum Computing, funded by European Union
NextGenerationEU, reference code CN{\textunderscore}00000013; M\'exico -- Consejo
Nacional de Ciencia y Tecnolog\'\i{}a (CONACYT) No.~167733; Universidad
Nacional Aut\'onoma de M\'exico (UNAM); PAPIIT DGAPA-UNAM; The Netherlands
-- Ministry of Education, Culture and Science; Netherlands Organisation
for Scientific Research (NWO); Dutch national e-infrastructure with the
support of SURF Cooperative; Poland -- Ministry of Education and
Science, grants No.~DIR/WK/2018/11 and 2022/WK/12; National Science
Centre, grants No.~2016/22/M/ST9/00198, 2016/23/B/ST9/01635,
2020/39/B/ST9/01398, and 2022/45/B/ST9/02163; Portugal -- Portuguese
national funds and FEDER funds within Programa Operacional Factores de
Competitividade through Funda\c{c}\~ao para a Ci\^encia e a Tecnologia
(COMPETE); Romania -- Ministry of Research, Innovation and Digitization,
CNCS-UEFISCDI, contract no.~30N/2023 under Romanian National Core
Program LAPLAS VII, grant no.~PN 23 21 01 02 and project number
PN-III-P1-1.1-TE-2021-0924/TE57/2022, within PNCDI III; Slovenia --
Slovenian Research Agency, grants P1-0031, P1-0385, I0-0033, N1-0111;
Spain -- Ministerio de Econom\'\i{}a, Industria y Competitividad
(FPA2017-85114-P and PID2019-104676GB-C32), Xunta de Galicia (ED431C
2017/07), Junta de Andaluc\'\i{}a (SOMM17/6104/UGR, P18-FR-4314) Feder Funds,
RENATA Red Nacional Tem\'atica de Astropart\'\i{}culas (FPA2015-68783-REDT) and
Mar\'\i{}a de Maeztu Unit of Excellence (MDM-2016-0692); USA -- Department of
Energy, Contracts No.~DE-AC02-07CH11359, No.~DE-FR02-04ER41300,
No.~DE-FG02-99ER41107 and No.~DE-SC0011689; National Science Foundation,
Grant No.~0450696; The Grainger Foundation; Marie Curie-IRSES/EPLANET;
European Particle Physics Latin American Network; and UNESCO.
\end{sloppypar}

\end{document}